%% file: archetypes-and-gender-revtex4.tex
\DocumentMetadata{uncompress} 

\documentclass[
  pre,
  twocolumn,
  twoside,
  byrevtex,
  superscriptaddress,
  floatfix,
  longbibliography,
  nofootinbib,
]{revtex4-2}

\usepackage[realmainfile]{currfile}
\usepackage{multirow}
\input{\currfilebase.settings}
\input{\filenamebase.settings}
\input{\filenamebase.settings-archetypometrics}

\usepackage{ragged2e}

\raggedright

\usepackage[
  font=small,
  labelfont=bf,
  justification=raggedright,
  singlelinecheck=false
]{caption}

\setboolean{twocolswitch}{true}

\usepackage{ifluatex}
\usepackage{newpax}
\newpaxsetup{usefileattributes=true}

\ifluatex
   \directlua{
    require("newpax")
    newpax.writenewpax(
      "figures/masculine--feminine-\Ncharacters-\Ntraits-\Nstories"
    )
    newpax.writenewpax(
      "figures/feminine--masculine-\Ncharacters-\Ntraits-\Nstories"
    )
  }
\fi

\begin{document}

\title{\protect
\input{\filenamebase.title}
}

\input{\filenamebase.author}

\date{\today}

\begin{abstract}
  \protect
  \input{\filenamebase.abs} 
\end{abstract}

\maketitle

\setlength{\parskip}{1\baselineskip plus .1\baselineskip  minus .1\baselineskip}


\input{\filenamebase.body}


\bigskip

\acknowledgments
\input{\filenamebase.acknowledgments}

\input{\filenamebase.biblio}









\newwrite\tempfile
\immediate\openout\tempfile=startsupp.txt
\immediate\write\tempfile{\thepage}
\immediate\closeout\tempfile

\setcounter{page}{1}
\renewcommand{\thepage}{S\arabic{page}}
 
\renewcommand{\thefigure}{S\arabic{figure}}
\setcounter{figure}{0}
 
\renewcommand{\thetable}{S\arabic{table}}
\setcounter{table}{0}
 
\renewcommand{\thesection}{S\arabic{section}}
\setcounter{section}{0}

\input{\filenamebase.supplementary}

\end{document}

%% file: archetypes-and-gender.settings.tex
\lefthyphenmin=3
\righthyphenmin=2

\usepackage[table]{xcolor}




\usepackage{xurl}
\urlstyle{same}

\PassOptionsToPackage{hyphens}{url}\usepackage{hyperref}
\makeatletter
\g@addto@macro{\UrlBreaks}{\UrlOrds}
\makeatother

\usepackage{colortbl}

\definecolor{goodblue}{RGB}{0, 91, 187}

\usepackage{hyperref}
\hypersetup{
  hypertexnames=false,
  colorlinks=true,
  allcolors=goodblue,
  urlcolor=.,
  linkcolor=goodblue,
  citecolor=goodblue,
  pdfborder={0 0 0},
  breaklinks=true,
  pdfcreator = {},  
  pdfproducer = {}  
}





\usepackage{stmaryrd}


\newcommand{\externallinksymbol}{{\tiny$^{{}_{\nnearrow}}$\!\!}}
\newcommand{\internallinksymbol}{{$^{\Rsh}$\!\!}}
\newcommand{\paperlinksymbol}{\externallinksymbol}



\usepackage [english]{babel}
\usepackage [autostyle, english = american]{csquotes}
\MakeOuterQuote{"}

\usepackage{listings}

\usepackage[normalem]{ulem}


\usepackage{colortbl}

\makeatletter

\def\CT@@do@color{%
  \global\let\CT@do@color\relax
  \@tempdima\wd\z@
  \advance\@tempdima\@tempdimb
  \advance\@tempdima\@tempdimc
  \advance\@tempdimb\tabcolsep
  \advance\@tempdimc\tabcolsep
  \advance\@tempdima2\tabcolsep
  \kern-\@tempdimb
  \leaders\vrule
  \hskip\@tempdima\@plus  1fill
  \kern-\@tempdimc
  \hskip-\wd\z@ \@plus -1fill }
\makeatother


\usepackage{xcolor}

\definecolor{olivegreen}{rgb}{0.33333,.41961,0.18431}
\definecolor{forestgreen}{rgb}{0.13333,.5451,0.13333}

\definecolor{lightgrey}{rgb}{0.7,0.7,0.7}
\definecolor{verylightgrey}{rgb}{0.90,0.90,0.90}
\definecolor{veryverylightgrey}{rgb}{0.95,0.95,0.95}
\definecolor{grey}{rgb}{0.5,0.5,0.5}
\definecolor{darkgrey}{rgb}{0.3,0.3,0.3}
\definecolor{verydarkgrey}{rgb}{0.15,0.15,0.15}

\usepackage{soul} 


\definecolor{headerblue}{HTML}{33367E}
\definecolor{unitednationsblue}{HTML}{4D88FF}

\definecolor{charcoal}{HTML}{36454F}
\definecolor{cinerous}{HTML}{98817B}
\definecolor{feldgrau}{HTML}{4D5D53}
\definecolor{glaucous}{HTML}{6082B6}
\definecolor{arsenic}{HTML}{3B444B}
\definecolor{xanadu}{HTML}{738678}

\definecolor{firebrick}{HTML}{B22222}
\definecolor{orangered}{HTML}{FF4500}
\definecolor{tomato}{HTML}{FF6347}

\definecolor{purpletaupe}{HTML}{3B444B}


\definecolor{headerorange}{RGB}{255,116,0}
\definecolor{headergray}{RGB}{230,230,230}

\definecolor{headerpop}{RGB}{230,230,230}

\definecolor{magmalight}{RGB}{252,251,195}
\definecolor{magmalightalt}{RGB}{250,240,184}
\definecolor{magmamedium}{RGB}{245,200,146}
\definecolor{magmadark}{RGB}{224,106,98}

\definecolor{icelight}{RGB}{223,242,244}
\definecolor{icelightalt}{RGB}{189,222,226}
\definecolor{icemedium}{RGB}{132,184,204}
\definecolor{icedark}{RGB}{103,153,191}

\definecolor{traitrowcolor}{RGB}{223,242,244}
\definecolor{traitrowcoloralt}{RGB}{189,222,226}

\definecolor{characterrowcolor}{RGB}{252,251,195}
\definecolor{characterrowcoloralt}{RGB}{250,240,184}

\definecolor{archetyperowcolor}{RGB}{255,213,212} 
\definecolor{archetyperowcoloralt}{RGB}{255,182,179} 

\definecolor{datasetrowcolor}{RGB}{232,244,234}
\definecolor{datasetrowcoloralt}{RGB}{210,231,214}

\definecolor{todoblue}{RGB}{0, 91, 187}

\usepackage{graphicx,epsfig,verbatim,enumerate}
\usepackage{amssymb,amsmath}
\usepackage{ifthen}

\usepackage{mathtools}


\newboolean{twocolswitch}



\newcommand{\sindex}[1]{}
\newcommand{\nindex}[1]{}

\newcommand{\www}[1]{\url{#1}}

\usepackage{lettrine}







%% file: archetypes-and-gender.settings-archetypometrics.tex

\newcommand{\Ncharacters}{2000}
\newcommand{\Ntraits}{464}
\newcommand{\Nstories}{341}

\newcommand{\Ncharactersbase}{2000}

\newcommand{\Ncharactersmain}{2000}
\newcommand{\Ntraitsmain}{464}
\newcommand{\Nstoriesmain}{341}

\newcommand{\Ncharactersmainone}{0800}
\newcommand{\Ntraitsmainone}{235}
\newcommand{\Nstoriesmainone}{090}

\newcommand{\Ncharactersmaintwo}{1600}
\newcommand{\Ntraitsmaintwo}{364}
\newcommand{\Nstoriesmaintwo}{241}


\newcommand{\onlinesiteshort}{compstorylab.org/archetypometrics}
\newcommand{\onlinesite}{https://\onlinesiteshort}


\newcommand{\semdiffsign}{\Leftrightarrow}
\newcommand{\semdiffsignleft}{\Leftarrow}
\newcommand{\semdiffsignright}{\Rightarrow}

\newcommand{\semdiff}[2]{\{#1\,$\semdiffsign$\,#2\}}

\newcommand{\semdiffright}[2]{\{#1\,$\semdiffsignright$\,#2\}}

\newcommand{\semdiffbold}[2]{\{\textbf{#1}\,$\semdiffsign$\,\textbf{#2}\}}
\newcommand{\semdiffboldleft}[2]{\{\textbf{#1}\,$\semdiffsignleft$\,#2\}}
\newcommand{\semdiffboldright}[2]{\{#1\,$\semdiffsignright$\,\textbf{#2}\}}

\newcommand{\semdiffmath}[2]{\{\textnormal{\textbf{#1}}\!\semdiffsign\!{\textnormal{\textbf{#2}}\}}}
\newcommand{\semdiffmathleft}[2]{\{\textnormal{\textbf{#1}}\!\semdiffsignleft\!{\textnormal{#2}\}}}
\newcommand{\semdiffmathright}[2]{\{\textnormal{#1}\!\semdiffsignright\!{\textnormal{\textbf{#2}}\}}}

\newcommand{\archetype}[1]{\archetypelinkbase{#1}}

\newcommand{\datasetsymbol}{\colorbox{datasetrowcolor}{\textcolor{black}{\stackanchor{\scalebox{\babaisyouboxscale}{DA}}{\scalebox{\babaisyouboxscale}{TA}}}}}

\newcommand{\dataset}[1]{\mbox{\datasetsymbol}_{#1}}


\newcommand{\datasetbase}[1]{
    \IfEqCase{#1}{
        {0800}{\datasetsymbol{1}}
        {1600}{\datasetsymbol{2}}
        {2000}{\datasetsymbol{3}}
    }[\PackageError{datasetbase}{Undefined option to datasetbase: #1}{}]%
}%

\newcommand{\datasetNcharacters}[1]{
    \IfEqCase{#1}{
        {1}{800}
        {2}{1600}
        {3}{2000}
    }[\PackageError{datasetNcharacters}{Undefined option to datasetNcharacters: #1}{}]%
}%

\newcommand{\datasetNtraits}[1]{
    \IfEqCase{#1}{
        {1}{235}
        {2}{364}
        {3}{464}
    }[\PackageError{datasetNtraits}{Undefined option to datasetNtraits: #1}{}]%
}%

\newcommand{\datasetNstories}[1]{
    \IfEqCase{#1}{
        {1}{90}
        {2}{241}
        {3}{341}
    }[\PackageError{datasetNstories}{Undefined option to datasetNstories: #1}{}]%
}%






\newcommand{\padzero}[1]{\ifnum #1 < 10 0\fi #1}


\usepackage{siunitx}


\newcommand\zeropad[2]{%
  \ifnum#2<0\relax%
    {\ensuremath-}\zeropadA{#1}{\the\numexpr#2*-1\relax}%
  \else%
    \zeropadA{#1}{#2}%
  \fi%
}
\def\zeropadA#1#2{%
  \ifnum1#2<1#1
    \zeropadA{#1}{0#2}%
  \else%
    #2%
  \fi%
}



\usepackage{xstring}



\newcommand{\archetypesemdiff}[1]{
    \IfEqCase{#1}{
        {1}{\semdiffbold{\archetypelinkbase{Fool}}{\archetypelinkbase{Hero}}}                  
        {2}{\semdiffbold{\archetypelinkbase{Angel}}{\archetypelinkbase{Demon}}}                
        {3}{\semdiffbold{\archetypelinkbase{Traditionalist}}{\archetypelinkbase{Adventurer}}}  
        {4}{\semdiffbold{\archetypelinksimple{Lone-Wolf}{Lone~Wolf}}{\archetypelinkbase{Diva}}}
        {5}{\semdiffbold{\archetypelinkbase{Outcast}}{\archetypelinkbase{Sophisticate}}}       
        {6}{\semdiffbold{\archetypelinkbase{Brute}}{\archetypelinkbase{Geek}}}                 
    }[\PackageError{archetypesemdiff}{Undefined option to archetypesemdiff: #1}{}]%
}%

\newcommand{\archetypesemdiffleft}[1]{
    \IfEqCase{#1}{
        {1}{\semdiffboldleft{\archetypelinkbase{Fool}}{\archetypelinkbase{Hero}}}                  
        {2}{\semdiffboldleft{\archetypelinkbase{Angel}}{\archetypelinkbase{Demon}}}                
        {3}{\semdiffboldleft{\archetypelinkbase{Traditionalist}}{\archetypelinkbase{Adventurer}}}  
        {4}{\semdiffboldleft{\archetypelinksimple{Lone-Wolf}{Lone~Wolf}}{\archetypelinkbase{Diva}}}
        {5}{\semdiffboldleft{\archetypelinkbase{Outcast}}{\archetypelinkbase{Sophisticate}}}       
        {6}{\semdiffboldleft{\archetypelinkbase{Brute}}{\archetypelinkbase{Geek}}}                 
    }[\PackageError{archetypesemdiff}{Undefined option to archetypesemdiff: #1}{}]%
}%

\newcommand{\archetypesemdiffright}[1]{
    \IfEqCase{#1}{
        {1}{\semdiffboldright{\archetypelinkbase{Fool}}{\archetypelinkbase{Hero}}}                  
        {2}{\semdiffboldright{\archetypelinkbase{Angel}}{\archetypelinkbase{Demon}}}                
        {3}{\semdiffboldright{\archetypelinkbase{Traditionalist}}{\archetypelinkbase{Adventurer}}}  
        {4}{\semdiffboldright{\archetypelinksimple{Lone-Wolf}{Lone~Wolf}}{\archetypelinkbase{Diva}}}
        {5}{\semdiffboldright{\archetypelinkbase{Outcast}}{\archetypelinkbase{Sophisticate}}}       
        {6}{\semdiffboldright{\archetypelinkbase{Brute}}{\archetypelinkbase{Geek}}}                 
    }[\PackageError{archetypesemdiff}{Undefined option to archetypesemdiff: #1}{}]%
}%

\newcommand{\archetypesemdiffmath}[1]{
  \IfEqCase{#1}{
    {1}{\semdiffmath{\archetypelinkbase{Fool}}{\archetypelinkbase{Hero}}}
    {2}{\semdiffmath{\archetypelinkbase{Angel}}{\archetypelinkbase{Demon}}}
    {3}{\semdiffmath{\archetypelinkbase{Traditionalist}}{\archetypelinkbase{Adventurer}}}
    {4}{\semdiffmath{\archetypelinksimple{Lone-Wolf}{Lone~Wolf}}{\archetypelinkbase{Diva}}}
    {5}{\semdiffmath{\archetypelinkbase{Outcast}}{\archetypelinkbase{Sophisticate}}}
    {6}{\semdiffmath{\archetypelinkbase{Brute}}{\archetypelinkbase{Geek}}}                 
  }[\PackageError{archetypesemdiffmath}{Undefined option to archetypesemdiffmath: #1}{}]%
}%

\newcommand{\archetypesemdiffmathswap}[1]{
  \IfEqCase{#1}{
    {1}{\semdiffmath{\archetypelinkbase{Hero}}{\archetypelinkbase{Fool}}}
    {2}{\semdiffmath{\archetypelinkbase{Demon}}{\archetypelinkbase{Angel}}}
    {3}{\semdiffmath{\archetypelinkbase{Adventurer}}{\archetypelinkbase{Traditionalist}}}
    {4}{\semdiffmath{\archetypelinkbase{Diva}}{\archetypelinksimple{Lone-Wolf}{Lone~Wolf}}}
    {5}{\semdiffmath{\archetypelinkbase{Sophisticate}}{\archetypelinkbase{Outcast}}}
    {6}{\semdiffmath{{\archetypelinkbase{Geek}}\archetypelinkbase{Brute}}}
  }[\PackageError{archetypesemdiffmathswap}{Undefined option to archetypesemdiffmathswap: #1}{}]%
}%

\newcommand{\archetypesemdiffmathleft}[1]{
  \IfEqCase{#1}{
    {1}{\semdiffmathleft{\archetypelinkbase{Fool}}{\archetypelinkbase{Hero}}}                  
    {2}{\semdiffmathleft{\archetypelinkbase{Angel}}{\archetypelinkbase{Demon}}}                
    {3}{\semdiffmathleft{\archetypelinkbase{Traditionalist}}{\archetypelinkbase{Adventurer}}}  
    {4}{\semdiffmathleft{\archetypelinksimple{Lone-Wolf}{Lone~Wolf}}{\archetypelinkbase{Diva}}}
    {5}{\semdiffmathleft{\archetypelinkbase{Outcast}}{\archetypelinkbase{Sophisticate}}}       
    {6}{\semdiffmathleft{\archetypelinkbase{Brute}}{\archetypelinkbase{Geek}}}                 
  }[\PackageError{archetypesemdiffmathleft}{Undefined option to archetypesemdiffmathleft: #1}{}]%
}%

\newcommand{\archetypesemdiffmathright}[1]{
  \IfEqCase{#1}{
    {1}{\semdiffmathright{\archetypelinkbase{Fool}}{\archetypelinkbase{Hero}}}                  
    {2}{\semdiffmathright{\archetypelinkbase{Angel}}{\archetypelinkbase{Demon}}}                
    {3}{\semdiffmathright{\archetypelinkbase{Traditionalist}}{\archetypelinkbase{Adventurer}}}  
    {4}{\semdiffmathright{\archetypelinksimple{Lone-Wolf}{Lone~Wolf}}{\archetypelinkbase{Diva}}}
    {5}{\semdiffmathright{\archetypelinkbase{Outcast}}{\archetypelinkbase{Sophisticate}}}       
    {6}{\semdiffmathright{\archetypelinkbase{Brute}}{\archetypelinkbase{Geek}}}                 
  }[\PackageError{archetypesemdiffmathright}{Undefined option to archetypesemdiffmathright: #1}{}]%
}%

\newcommand{\essentialsemdiff}[1]{
  \IfEqCase{#1}{
    {1}{\semdiff{\essentialtraitlinknegative{1}{weak/incompetent/lazy/stupid}}{\essentialtraitlinkpositive{1}{powerful/capable/purposeful/intelligent}}}
    {2}{\semdiff{\essentialtraitlinknegative{2}{safe/pure/virtuous/humble}}{\essentialtraitlinkpositive{2}{dangerous/depraved/corrupt/arrogant}}}
    {3}{\semdiff{\essentialtraitlinknegative{3}{serious/predictable/humorless/uncreative}}{\essentialtraitlinkpositive{3}{playful/unpredictable/funny/creative}}}
    {4}{\semdiff{\essentialtraitlinknegative{4}{rugged/stoic/independent/blunt}}{\essentialtraitlinkpositive{4}{refined/dramatic/dependent/sensitive}}}
    {5}{\semdiff{\essentialtraitlinknegative{5}{unlucky/unsophisticated/traumatized}}{\essentialtraitlinkpositive{5}{fortunate/sophisticated/confident}}}
    {6}{\semdiff{\essentialtraitlinknegative{6}{physical/mainstream/simple-minded}}{\essentialtraitlinkpositive{6}{intellectual/weird/complex}}}
    {7}{\semdiff{\essentialtraitlinknegative{7}{dramatic/attractive/young}}{\essentialtraitlinkpositive{7}{comedic/ugly/old}}}
    {8}{\semdiff{\essentialtraitlinknegative{8}{spiritual/rural/historical}}{\essentialtraitlinkpositive{8}{skeptical/urban/modern}}}
    {9}{\semdiff{\essentialtraitlinknegative{9}{old/historical/low-tempo}}{\essentialtraitlinkpositive{9}{young/modern/high-tempo}}}
    {10}{\semdiff{\essentialtraitlinknegative{10}{feminine/luddite}}{\essentialtraitlinkpositive{10}{masculine/technophile}}}
    {11}{\semdiff{\essentialtraitlinknegative{11}{secondary/street-wise}}{\essentialtraitlinkpositive{11}{primary/sheltered}}}
  }[\PackageError{essentialsemdiff}{Undefined option to essentialsemdiff: #1}{}]%
}%

\newcommand{\essentialsemdiffloose}[1]{
  \IfEqCase{#1}{
    {1}{\semdiff{\essentialtraitlinknegative{1}{weak, incompetent, lazy, stupid}}{\essentialtraitlinkpositive{1}{powerful, capable, purposeful, intelligent}}}
    {2}{\semdiff{\essentialtraitlinknegative{2}{safe, pure, virtuous, humble}}{\essentialtraitlinkpositive{2}{dangerous, depraved, corrupt, arrogant}}}
    {3}{\semdiff{\essentialtraitlinknegative{3}{serious, predictable, humorless, uncreative}}{\essentialtraitlinkpositive{3}{playful, unpredictable, funny, creative}}}
    {4}{\semdiff{\essentialtraitlinknegative{4}{rugged, stoic, independent, blunt}}{\essentialtraitlinkpositive{4}{refined, dramatic, dependent, sensitive}}}
    {5}{\semdiff{\essentialtraitlinknegative{5}{unlucky, unsophisticated, traumatized}}{\essentialtraitlinkpositive{5}{fortunate, sophisticated, confident}}}
    {6}{\semdiff{\essentialtraitlinknegative{6}{physical, mainstream, simple-minded}}{\essentialtraitlinkpositive{6}{intellectual, weird, complex}}}
    {7}{\semdiff{\essentialtraitlinknegative{7}{dramatic, attractive, young}}{\essentialtraitlinkpositive{7}{comedic, ugly, old}}}
    {8}{\semdiff{\essentialtraitlinknegative{8}{spiritual, rural, historical}}{\essentialtraitlinkpositive{8}{skeptical, urban, modern}}}
    {9}{\semdiff{\essentialtraitlinknegative{9}{old, historical, low-tempo}}{\essentialtraitlinkpositive{9}{young, modern, high-tempo}}}
    {10}{\semdiff{\essentialtraitlinknegative{10}{feminine, luddite}}{\essentialtraitlinkpositive{10}{masculine, technophile}}}
    {11}{\semdiff{\essentialtraitlinknegative{11}{secondary, street-wise}}{\essentialtraitlinkpositive{11}{primary, sheltered}}}
  }[\PackageError{essentialsemdiffloose}{Undefined option to essentialsemdiffloose: #1}{}]%
}%

\newcommand{\essentialsemdifflooseleft}[1]{
    \IfEqCase{#1}{
        {1}{\semdiff{\textbf{\essentialtraitlinknegative{1}{weak, incompetent, lazy, stupid}}}{\essentialtraitlinkpositive{1}{powerful, capable, purposeful, intelligent}}}
        {2}{\semdiff{\textbf{\essentialtraitlinknegative{2}{safe, pure, virtuous, humble}}}{\essentialtraitlinkpositive{2}{dangerous, depraved, corrupt, arrogant}}}
        {3}{\semdiff{\textbf{\essentialtraitlinknegative{3}{serious, predictable, humorless, uncreative}}}{\essentialtraitlinkpositive{3}{playful, unpredictable, funny, creative}}}
        {4}{\semdiff{\textbf{\essentialtraitlinknegative{4}{rugged, stoic, independent, blunt}}}{\essentialtraitlinkpositive{4}{refined, dramatic, dependent, sensitive}}}
        {5}{\semdiff{\textbf{\essentialtraitlinknegative{5}{unlucky, unsophisticated, traumatized}}}{\essentialtraitlinkpositive{5}{fortunate, sophisticated, confident}}}
        {6}{\semdiff{\textbf{\essentialtraitlinknegative{6}{physical, mainstream, simple-minded}}}{\essentialtraitlinkpositive{6}{intellectual, weird, complex}}}
        {7}{\semdiff{\textbf{\essentialtraitlinknegative{7}{dramatic, attractive, young}}}{\essentialtraitlinkpositive{7}{comedic, ugly, old}}}
        {8}{\semdiff{\textbf{\essentialtraitlinknegative{8}{spiritual, rural, historical}}}{\essentialtraitlinkpositive{8}{skeptical, urban, modern}}}
        {9}{\semdiff{\textbf{\essentialtraitlinknegative{9}{old, historical, low-tempo}}}{\essentialtraitlinkpositive{9}{young, modern, high-tempo}}}
        {10}{\semdiff{\textbf{\essentialtraitlinknegative{10}{feminine, luddite}}}{\essentialtraitlinkpositive{10}{masculine, technophile}}}
        {11}{\semdiff{\textbf{\essentialtraitlinknegative{11}{secondary, street-wise}}}{\essentialtraitlinkpositive{11}{primary, sheltered}}}
    }[\PackageError{essentialsemdifflooseleft}{Undefined option to essentialsemdifflooseleft: #1}{}]%
 }%

\newcommand{\essentialsemdifflooseright}[1]{
  \IfEqCase{#1}{
    {1}{\semdiff{\essentialtraitlinknegative{1}{weak, incompetent, lazy, stupid}}{\textbf{\essentialtraitlinkpositive{1}{powerful, capable, purposeful, intelligent}}}}
    {2}{\semdiff{\essentialtraitlinknegative{2}{safe, pure, virtuous, humble}}{\textbf{\essentialtraitlinkpositive{2}{dangerous, depraved, corrupt, arrogant}}}}
    {3}{\semdiff{\essentialtraitlinknegative{3}{serious, predictable, humorless, uncreative}}{\textbf{\essentialtraitlinkpositive{3}{playful, unpredictable, funny, creative}}}}
    {4}{\semdiff{\essentialtraitlinknegative{4}{rugged, stoic, independent, blunt}}{\textbf{\essentialtraitlinkpositive{4}{refined, dramatic, dependent, sensitive}}}}
    {5}{\semdiff{\essentialtraitlinknegative{5}{unlucky, unsophisticated, traumatized}}{\textbf{\essentialtraitlinkpositive{5}{fortunate, sophisticated, confident}}}}
    {6}{\semdiff{\essentialtraitlinknegative{6}{physical, mainstream, simple-minded}}{\textbf{\essentialtraitlinkpositive{6}{intellectual, weird, complex}}}}
    {7}{\semdiff{\essentialtraitlinknegative{7}{dramatic, attractive, young}}{\textbf{\essentialtraitlinkpositive{7}{comedic, ugly, old}}}}
    {8}{\semdiff{\essentialtraitlinknegative{8}{spiritual, historical, rural}}{\textbf{\essentialtraitlinkpositive{8}{skeptical, urban, modern}}}}
    {9}{\semdiff{\essentialtraitlinknegative{9}{old, historical, low-tempo}}{\textbf{\essentialtraitlinkpositive{9}{young, modern, high-tempo}}}}
    {10}{\semdiff{\essentialtraitlinknegative{10}{feminine, luddite}}{\textbf{\essentialtraitlinkpositive{10}{masculine, technophile}}}}
    {11}{\semdiff{\essentialtraitlinknegative{11}{secondary, street-wise}}{\textbf{\essentialtraitlinkpositive{11}{primary, sheltered}}}}
  }[\PackageError{essentialsemdifflooseright}{Undefined option to essentialsemdifflooseright: #1}{}]%
}%

\newcommand{\essentialsemdiffmathleft}[1]{
    \IfEqCase{#1}{
        {1}{\semdiffmathleft{weak/incompetent/lazy/stupid}{powerful/capable/purposeful/intelligent}}
        {2}{\semdiffmathleft{safe/pure/virtuous/humble}{dangerous/depraved/corrupt/arrogant}}
        {3}{\semdiffmathleft{serious/predictable/humorless/uncreative}{playful/unpredictable/funny/creative}}
        {4}{\semdiffmathleft{rugged/stoic/independent/blunt}{refined/dramatic/dependent/sensitive}}
        {5}{\semdiffmathleft{unlucky/unsophisticated/traumatized}{fortunate/sophisticated/confident}}
        {6}{\semdiffmathleft{physical/mainstream/simple-minded}{intellectual/weird/complex}}
        {7}{\semdiffmathleft{dramatic/attractive/young}{comedic/ugly/old}}
        {8}{\semdiffmathleft{spiritual/rural/historical}{skeptical/urban/modern}}
        {9}{\semdiffmathleft{old/historical/low-tempo}{young/modern/high-tempo}}
        {10}{\semdiffmathleft{feminine/luddite}{masculine/technophile}}
        {11}{\semdiffmathleft{secondary/street-wise}{primary/sheltered}}
    }[\PackageError{essentialsemdiffmathleft}{Undefined option to essentialsemdiffmathleft: #1}{}]%
}%

\newcommand{\essentialsemdiffmathright}[1]{
    \IfEqCase{#1}{
        {1}{\semdiffmathright{weak/incompetent/lazy/stupid}{powerful/capable/purposeful/intelligent}}
        {2}{\semdiffmathright{safe/pure/virtuous/humble}{dangerous/depraved/corrupt/arrogant}}
        {3}{\semdiffmathright{serious/predictable/humorless/uncreative}{playful/unpredictable/funny/creative}}
        {4}{\semdiffmathright{rugged/stoic/independent/blunt}{refined/dramatic/dependent/sensitive}}
        {5}{\semdiffmathright{unlucky/unsophisticated/traumatized}{fortunate/sophisticated/confident}}
        {6}{\semdiffmathright{physical/mainstream/simple-minded}{intellectual/weird/complex}}
        {7}{\semdiffmathright{dramatic/attractive/young}{comedic/ugly/old}}
        {8}{\semdiffmathright{spiritual/rural/historical}{skeptical/urban/modern}}
        {9}{\semdiffmathright{old/historical/low-tempo}{young/modern/high-tempo}}
        {10}{\semdiffmathright{feminine/luddite}{masculine/technophile}}
        {11}{\semdiffmathright{secondary/street-wise}{primary/sheltered}}
    }[\PackageError{essentialsemdiffmathright}{Undefined option to essentialsemdiffmathright: #1}{}]%
}%

\newcommand{\essentialsemdiffmath}[1]{
    \IfEqCase{#1}{
        {1}{\semdiffmath{weak/incompetent/lazy/stupid}{powerful/capable/purposeful/intelligent}}
        {2}{\semdiffmath{safe/pure/virtuous/humble}{dangerous/depraved/corrupt/arrogant}}
        {3}{\semdiffmath{serious/predictable/humorless/uncreative}{playful/unpredictable/funny/creative}}
        {4}{\semdiffmath{rugged/stoic/independent/blunt}{refined/dramatic/dependent/sensitive}}
        {5}{\semdiffmath{unlucky/unsophisticated/traumatized}{fortunate/sophisticated/confident}}
        {6}{\semdiffmath{physical/mainstream/simple-minded}{intellectual/weird/complex}}
        {7}{\semdiffmath{dramatic/attractive/young}{comedic/ugly/old}}
        {8}{\semdiffmath{spiritual/rural/historical}{skeptical/urban/modern}}
        {9}{\semdiffmath{old/historical/low-tempo}{young/modern/high-tempo}}
        {10}{\semdiffmath{feminine/luddite}{masculine/technophile}}
        {11}{\semdiffmath{secondary/street-wise}{primary/sheltered}}
    }[\PackageError{essentialsemdiffmath}{Undefined option to essentialsemdiffmath: #1}{}]%
}%

\newcommand{\ousiometricsemdiff}[1]{
    \IfEqCase{#1}{
        {1}{\semdiffbold{weak}{powerful}}
        {2}{\semdiffbold{safe}{dangerous}}
        {3}{\semdiffbold{structured}{unstructured}}
    }[\PackageError{ousiometricsemdiff}{Undefined option to ousiometricsemdiff: #1}{}]%
}%

\newcommand{\ousiometricsemdiffmath}[1]{
    \IfEqCase{#1}{
        {1}{\semdiffmath{weak}{powerful}}
        {2}{\semdiffmath{safe}{dangerous}}
        {3}{\semdiffmath{structured}{unstructured}}
    }[\PackageError{ousiometricsemdiffmath}{Undefined option to ousiometricsemdiffmath: #1}{}]%
}%

\newcommand{\ousiometricsemdiffmathleft}[1]{
    \IfEqCase{#1}{
        {1}{\semdiffmathleft{weak}{powerful}}
        {2}{\semdiffmathleft{safe}{dangerous}}
        {3}{\semdiffmathleft{structured}{unstructured}}
    }[\PackageError{ousiometricsemdiffmathleft}{Undefined option to ousiometricsemdiffmathleft: #1}{}]%
}%

\newcommand{\ousiometricsemdiffmathright}[1]{
    \IfEqCase{#1}{
        {1}{\semdiffmathright{weak}{powerful}}
        {2}{\semdiffmathright{safe}{dangerous}}
        {3}{\semdiffmathright{structured}{unstructured}}
    }[\PackageError{ousiometricsemdiffmathright}{Undefined option to ousiometricsemdiffmathright: #1}{}]%
}%

\newcommand{\dimensiontype}[1]{
    \IfEqCase{#1}{
        {1}{Primary archetype}
        {2}{Primary archetype}
        {3}{Primary archetype}
        {4}{Secondary Archetype}
        {5}{Secondary Archetype}
        {6}{Secondary Archetype}
        {7}{Complex Trait}
        {8}{Complex Trait}
        {9}{Complex Trait}
        {10}{Complex Trait}
        {11}{Complex Trait}
    }[\PackageError{dimensiontype}{Undefined option to dimensiontype: #1}{}]%
}%




\usepackage{stackengine}
\setstackgap{S}{1pt}
\setlength{\fboxsep}{1pt}

\newcommand{\babaisyouboxscale}{0.48}

\newcommand{\archetyperatiosymbol}[2]{{\colorbox{#1}{\textcolor{#2}{\stackanchor{\scalebox{\babaisyouboxscale}{AR}}{\scalebox{\babaisyouboxscale}{CH}}}}}}

\newcommand{\appendixsymbol}[2]{{\colorbox{#1}{\textcolor{#2}{\stackanchor{\scalebox{\babaisyouboxscale}{AP}}{\scalebox{\babaisyouboxscale}{DX}}}}}}


\newcommand{\archetypometricshome}{\onlinesite}
\newcommand{\cardsdir}{\archetypometricshome/cards}


\newcommand{\characterlinksimple}[2]{\href{\cardsdir/#1-\Ncharacters-\Ntraits-\Nstories.pdf}{\textcolor{verydarkgrey}{#2\paperlinksymbol}}}

\newcommand{\characterlinksimpledataset}[3]{
  \IfEqCase{#3}{
    {1}{\href{\cardsdir/#1-\Ncharactersmainone-\Ntraitsmainone-\Nstoriesmainone.pdf}{\textcolor{verydarkgrey}{#2\colorbox{datasetrowcolor}{$\dataset{#3}$}\paperlinksymbol}}}
    {2}{\href{\cardsdir/#1-\Ncharactersmaintwo-\Ntraitsmaintwo-\Nstoriesmaintwo.pdf}{\textcolor{verydarkgrey}{#2\colorbox{datasetrowcolor}{$\dataset{#3}$}\paperlinksymbol}}}
    {3}{\href{\cardsdir/#1-\Ncharactersmain-\Ntraitsmain-\Nstoriesmain.pdf}{\textcolor{verydarkgrey}{#2\colorbox{datasetrowcolor}{$\dataset{#3}$}\paperlinksymbol}}}
  }[\PackageError{characterlinksimpledataset}{Undefined option to characterlinksimpledataset: #1}{}]%
}


\newcommand{\traitlinksimple}[2]{\textcolor{verydarkgrey}{\semdiff{\href{\cardsdir/#2--#1-\Ncharacters-\Ntraits-\Nstories.pdf}{#1\paperlinksymbol}}{\href{\cardsdir/#1--#2-\Ncharacters-\Ntraits-\Nstories.pdf}{#2\paperlinksymbol}}}}

\newcommand{\traitlinksimpledataset}[3]{
  \IfEqCase{#3}{
    {1}{\textcolor{verydarkgrey}{\semdiff{\href{\cardsdir/#2--#1-\Ncharactersmainone-\Ntraitsmainone-\Nstoriesmainone.pdf}{#1\colorbox{datasetrowcolor}{$\dataset{#3}$}\paperlinksymbol}}{\href{\cardsdir/#1--#2-\Ncharactersmainone-\Ntraitsmainone-\Nstoriesmainone.pdf}{#2\colorbox{datasetrowcolor}{$\dataset{#3}$}\paperlinksymbol}}}}
    {2}{\textcolor{verydarkgrey}{\semdiff{\href{\cardsdir/#2--#1-\Ncharactersmaintwo-\Ntraitsmaintwo-\Nstoriesmaintwo.pdf}{#1\colorbox{datasetrowcolor}{$\dataset{#3}$}\paperlinksymbol}}{\href{\cardsdir/#1--#2-\Ncharactersmaintwo-\Ntraitsmaintwo-\Nstoriesmaintwo.pdf}{#2\colorbox{datasetrowcolor}{$\dataset{#3}$}\paperlinksymbol}}}}
    {3}{\textcolor{verydarkgrey}{\semdiff{\href{\cardsdir/#2--#1-\Ncharactersmain-\Ntraitsmain-\Nstoriesmain.pdf}{#1\colorbox{datasetrowcolor}{$\dataset{#3}$}\paperlinksymbol}}{\href{\cardsdir/#1--#2-\Ncharactersmain-\Ntraitsmain-\Nstoriesmain.pdf}{#2\colorbox{datasetrowcolor}{$\dataset{#3}$}\paperlinksymbol}}}}
  }[\PackageError{traitlinksimpledataset}{Undefined option to traitlinksimpledataset: #1}{}]%
}

\newcommand{\traitlinksimpleright}[2]{\href{\cardsdir/#1--#2-\Ncharacters-\Ntraits-\Nstories.pdf}{\textcolor{verydarkgrey}{\semdiffright{#1}{\textbf{#2}}\paperlinksymbol}}}

\newcommand{\traitlinksimpledatasetalt}[5]{
  \IfEqCase{#5}{
    {1}{\textcolor{verydarkgrey}{\semdiff{\href{\cardsdir/#2--#1-\Ncharactersmainone-\Ntraitsmainone-\Nstoriesmainone.pdf}{#3\colorbox{datasetrowcolor}{$\dataset{#5}$}\paperlinksymbol}}{\href{\cardsdir/#1--#2-\Ncharactersmainone-\Ntraitsmainone-\Nstoriesmainone.pdf}{#4\colorbox{datasetrowcolor}{$\dataset{#5}$}\paperlinksymbol}}}}
    {2}{\textcolor{verydarkgrey}{\semdiff{\href{\cardsdir/#2--#1-\Ncharactersmaintwo-\Ntraitsmaintwo-\Nstoriesmaintwo.pdf}{#3\colorbox{datasetrowcolor}{$\dataset{#5}$}\paperlinksymbol}}{\href{\cardsdir/#1--#2-\Ncharactersmaintwo-\Ntraitsmaintwo-\Nstoriesmaintwo.pdf}{#4\colorbox{datasetrowcolor}{$\dataset{#5}$}\paperlinksymbol}}}}
    {3}{\textcolor{verydarkgrey}{\semdiff{\href{\cardsdir/#2--#1-\Ncharactersmain-\Ntraitsmain-\Nstoriesmain.pdf}{#3\colorbox{datasetrowcolor}{$\dataset{#5}$}\paperlinksymbol}}{\href{\cardsdir/#1--#2-\Ncharactersmain-\Ntraitsmain-\Nstoriesmain.pdf}{#4\colorbox{datasetrowcolor}{$\dataset{#5}$}\paperlinksymbol}}}}
  }[\PackageError{traitlinksimpledatasetalt}{Undefined option to traitlinksimpledatasetalt: #1}{}]%
}


\newcommand{\storylinksimple}[2]{\href{\cardsdir/#1-\Ncharacters-\Ntraits-\Nstories.pdf}{\textcolor{darkgrey}{#2\paperlinksymbol}}}

\newcommand{\storylinksimpledataset}[3]{
  \IfEqCase{#3}{
    {1}{\href{\cardsdir/#1-\Ncharactersmainone-\Ntraitsmainone-\Nstoriesmainone.pdf}{\textcolor{verydarkgrey}{#2\colorbox{datasetrowcolor}{$\dataset{#3}$}\paperlinksymbol}}}
    {2}{\href{\cardsdir/#1-\Ncharactersmaintwo-\Ntraitsmaintwo-\Nstoriesmaintwo.pdf}{\textcolor{verydarkgrey}{#2\colorbox{datasetrowcolor}{$\dataset{#3}$}\paperlinksymbol}}}
    {3}{\href{\cardsdir/#1-\Ncharactersmain-\Ntraitsmain-\Nstoriesmain.pdf}{\textcolor{verydarkgrey}{#2\colorbox{datasetrowcolor}{$\dataset{#3}$}\paperlinksymbol}}}
  }[\PackageError{storylinksimpledataset}{Undefined option to storylinksimpledataset: #1}{}]%
}


\newcommand{\grouplinksimpledataset}[3]{
  \IfEqCase{#3}{
    {1}{\href{\cardsdir/#1-\Ncharactersmainone-\Ntraitsmainone-\Nstoriesmainone.pdf}{\textcolor{verydarkgrey}{#2\colorbox{datasetrowcolor}{$\dataset{#3}$}\paperlinksymbol}}}
    {2}{\href{\cardsdir/#1-\Ncharactersmaintwo-\Ntraitsmaintwo-\Nstoriesmaintwo.pdf}{\textcolor{verydarkgrey}{#2\colorbox{datasetrowcolor}{$\dataset{#3}$}\paperlinksymbol}}}
    {3}{\href{\cardsdir/#1-\Ncharactersmain-\Ntraitsmain-\Nstoriesmain.pdf}{\textcolor{verydarkgrey}{#2\colorbox{datasetrowcolor}{$\dataset{#3}$}\paperlinksymbol}}}
  }[\PackageError{grouplinksimpledataset}{Undefined option to grouplinksimpledataset: #1}{}]%
}



\newcommand{\archetypelinkbase}[1]{\href{\cardsdir/Archetype-#1-component-size-\Ncharacters-\Ntraits-\Nstories.pdf}{\textcolor{verydarkgrey}{#1\paperlinksymbol}}}

\newcommand{\archetypelinksimple}[2]{\href{\cardsdir/Archetype-#1-component-size-\Ncharacters-\Ntraits-\Nstories.pdf}{\textcolor{verydarkgrey}{#2\paperlinksymbol}}}

\newcommand{\archetypelinksimpledataset}[3]{
  \IfEqCase{#3}{
    {1}{\href{\cardsdir/Archetype-#1-component-size-\Ncharactersmainone-\Ntraitsmainone-\Nstoriesmainone.pdf}{\textcolor{verydarkgrey}{#2\colorbox{datasetrowcolor}{$\dataset{#3}$}\paperlinksymbol}}}
    {2}{\href{\cardsdir/Archetype-#1-component-size-\Ncharactersmaintwo-\Ntraitsmaintwo-\Nstoriesmaintwo.pdf}{\textcolor{verydarkgrey}{#2\colorbox{datasetrowcolor}{$\dataset{#3}$}\paperlinksymbol}}}
    {3}{\href{\cardsdir/Archetype-#1-component-size-\Ncharactersmain-\Ntraitsmain-\Nstoriesmain.pdf}{\textcolor{verydarkgrey}{#2\colorbox{datasetrowcolor}{$\dataset{#3}$}\paperlinksymbol}}}
  }[\PackageError{archetypelinksimpledataset}{Undefined option to archetypelinksimpledataset: #1}{}]%
}

\newcommand{\archetypelinkratiosimpleappendix}[2]{{\hypersetup{allcolors=.}\hyperref[page:N\Ncharactersbase_archetypometrics.archetypeclass-#1]{#2\archetyperatiosymbol{archetyperowcolor}{black}\,\appendixsymbol{verydarkgrey}{white}\internallinksymbol}}}





\newcommand{\essentialtraitlinknegative}[2]{\href{\cardsdir/Essential-Trait-\zeropad{000}{#1}-negative-component-size-\Ncharacters-\Ntraits-\Nstories.pdf}{\textcolor{verydarkgrey}{#2\paperlinksymbol}}}

\newcommand{\essentialtraitlinkpositive}[2]{\href{\cardsdir/Essential-Trait-\zeropad{000}{#1}-positive-component-size-\Ncharacters-\Ntraits-\Nstories.pdf}{\textcolor{verydarkgrey}{#2\paperlinksymbol}}}


\newcommand{\characterlink}[1]{
  \IfEqCase{#1}{
    }[\PackageError{characterlink}{Undefined option to characterlink: #1}{}]%
}%

\newcommand{\characterlinkinsert}[2]{
  \IfEqCase{#1}{
    }[\PackageError{characterlinkinsert}{Undefined option to characterlinkinsert: #1}{}]%
}%


%% file: archetypes-and-gender.title.tex
Archetypes and gender in fiction: 
\\
A data-driven mapping of gender stereotypes in stories 















%% file: archetypes-and-gender.author.tex
\author{
\firstname{Calla Glavin}
\surname{Beauregard}
}

\email{calla.beauregard@uvm.edu}

\affiliation{
  Computational Story Lab,
  Vermont Complex Systems Institute,
  MassMutual Center of Excellence in Complex Systems and Data Science,
  University of Vermont,
  Burlington, VT 05405, USA.
  }

\author{
\firstname{Julia Witte}
\surname{Zimmerman}
}

\affiliation{
  Computational Story Lab,
  Vermont Complex Systems Institute,
  MassMutual Center of Excellence in Complex Systems and Data Science,
  University of Vermont,
  Burlington, VT 05405, USA.
  }

\author{
\firstname{}
\surname{Ashley~M.~A.~Fehr}
}

\affiliation{
  Computational Story Lab,
  Vermont Complex Systems Institute,
  MassMutual Center of Excellence in Complex Systems and Data Science,
  University of Vermont,
  Burlington, VT 05405, USA.
  }

\author{
\firstname{}
\surname{Timothy~R.~Tangherlini}
}

\affiliation{
  Department of Scandinavian,
  Folklore Program,
  School of Information,
  Berkeley Institute for Data Science,
  University~of~California,~Berkeley,~Berkeley,~CA~94720-1500,~USA
  }

\author{
\firstname{Christopher M.}
\surname{Danforth}
}

\affiliation{
  Computational Story Lab,
  Vermont Complex Systems Institute,
  MassMutual Center of Excellence in Complex Systems and Data Science,
  University of Vermont,
  Burlington, VT 05405, USA.
  }

\affiliation{
  Department of Mathematics \& Statistics,
  University of Vermont,
  Burlington, VT 05405, USA.
  }

\author{
  \firstname{Peter Sheridan}
  \surname{Dodds}
}

\email{peter.dodds@uvm.edu}

\affiliation{
  Computational Story Lab,
  Vermont Complex Systems Institute,
  MassMutual Center of Excellence in Complex Systems and Data Science,
  University of Vermont,
  Burlington, VT 05405, USA.
  }

\affiliation{
  Department of Computer Science,
  University of Vermont,
  Burlington, VT 05405, USA.
}

\affiliation{
  Santa Fe Institute,
  1399 Hyde Park Rd,
  Santa Fe,
  NM 87501,
  USA
}

%% file: archetypes-and-gender.abs.tex

Fictional character representations reflect social norms and biases.
For example,
women are relatively underrepresented in television and film, irrespective of genre,
and are frequently stereotyped in these media.
Here, we draw on a data-driven operationalization of archetypes---archetypometrics---to explore the characterization of 2,000 canonically male and female characters. 
From an overall space of six pairs of base archetypes,
we find that canonically female characters
tend more toward Hero, Adventurer, Diva, and Sophisticate archetypes, while male characters,
tend toward Fool, Traditionalist, Outcast, Brute and Outcast types.
However, overarching patterns by gender 
nevertheless sustain traditional stereotypes:
The seemingly positive heroic 
bias toward females
is undercut by 
heroic female characters being 
more masculine than other female characters.
We discuss the societal implications of skewed archetype representation by character gender. 



%% file: archetypes-and-gender.body.tex
\section{Introduction}
\label{sec:Introduction}
In this paper, we explore the relationship
between archetypes, canonical gender,
and the trait of 
\traitlinksimple{masculine}{feminine}
for fictional characters in literature, film, and movies. 
We build directly on our previous work~\cite{dodds2025archetypometrics} where we analyzed over 72M fan-sourced ratings for 2,000 characters in 341 stories across 464 semantic differential traits,
drawing on the Open-Source Psychometrics Project~\cite{openpsychometrics}.  
We determined that the 464 dimensional space reduced to a six-dimensional orthogonal space of
archetypes~\cite{dodds2025archetypometrics}.\footnote{The concept of archetype was originated by Jung~\cite{Jung1969},
but can be traced back to the early work of the Greek philosopher Theophrastus~\cite{Theophrastus_Characters_Rusten2003}.
Per Ref.~\cite{dodds2025archetypometrics}, our operationalization of archetypes quantitatively captures recurrent character types across many narrative theories. 
}
The six archetype pairs break down into 
three primary and three secondary archetypes.
In Table~\ref{tab:archetypometrics.shortsummary},
we list the six pairs along with complex semantic differentials to help anchor the reader's understanding, and also record the variance explained by each pair.


\begin{table*}[t!]

  \centering
  
  
  \setlength{\extrarowheight}{2.5pt}
  
  \rowcolors{2}{white}{white}
  \small
  \begin{tabular}{lc}
    \multicolumn{2}{l}{\textbf{\normalsize Primary essential dimensions:}}
    \smallskip
    \\
    \hline
    \rowcolor{headerpop}
    \#.
    Primary archetype pair $\sim$ differentials 
    &
    \% Var. Expl.
    \\
    \hline
    \rowcolor{archetyperowcoloralt}
    ~1.\archetypesemdiff{1}
    &
\input{inputs/localized/tabessential_archetypes200_basevarexpl_N\Ncharactersbase_dim01}\unskip\%
    \\
    \rowcolor{archetyperowcolor}
    \multicolumn{2}{l}{
    ~~~$\sim$\essentialsemdiff{1}
    }
    \\
    \rowcolor{archetyperowcoloralt}
    ~2.\archetypesemdiff{2}
    &   \input{inputs/localized/tabessential_archetypes200_basevarexpl_N\Ncharactersbase_dim02}\unskip\% 
    \\
    \rowcolor{archetyperowcolor}
    \multicolumn{2}{l}{
    ~~~$\sim$\essentialsemdiff{2}
    }
    \\    
    \rowcolor{archetyperowcoloralt}
    ~3.\archetypesemdiff{3}
    &
\input{inputs/localized/tabessential_archetypes200_basevarexpl_N\Ncharactersbase_dim03}\unskip\%
    \\
    \rowcolor{archetyperowcolor}
    \multicolumn{2}{l}{
    ~~~$\sim$\essentialsemdiff{3}    
    }   
    \\  
    \smallskip
    \\
    \multicolumn{2}{l}{\textbf{\normalsize Secondary essential dimensions:}}
    \\
    \hline
    \rowcolor{headerpop}
    \#. 
    Primary archetype pair $\sim$ differentials 
    &
    \% Var. Expl.
    \\
    \hline
    \rowcolor{archetyperowcolor}
    ~4.\archetypesemdiff{4}
    &
\input{inputs/localized/tabessential_archetypes200_basevarexpl_N\Ncharactersbase_dim04}\unskip\%
    \\
    \rowcolor{archetyperowcolor}
    \multicolumn{2}{l}{
    ~~~$\sim$\essentialsemdiff{4}
    }
    \\
    \rowcolor{archetyperowcoloralt}
    ~5.\archetypesemdiff{5}
    &   \input{inputs/localized/tabessential_archetypes200_basevarexpl_N\Ncharactersbase_dim05}\unskip\% 
    \\
    \rowcolor{archetyperowcolor}
    \multicolumn{2}{l}{
    ~~~$\sim$\essentialsemdiff{5}
    }
    \\
    \rowcolor{archetyperowcoloralt}
    ~6.\archetypesemdiff{6}
    &
\input{inputs/localized/tabessential_archetypes200_basevarexpl_N\Ncharactersbase_dim06}\unskip\%
    \\
\rowcolor{archetyperowcolor}
\multicolumn{2}{l}{~~~$\sim$\essentialsemdiff{6}} 
    \\    
  \end{tabular}
  \caption{
    Primary and secondary archetype pairs listed with the semantic differential trait pairings that best capture the variance between different archetypes. Semantic differential trait pairs displayed from negative to positive values from left to right. Negative or positive values for traits denote directionality in each pairing; magnitude denotes strength of that specific trait. 
  }
  \label{tab:archetypometrics.shortsummary}
\end{table*}


Through archetypes, we can explore how the representation of fictional characters reflect or refract social norms, values, and beliefs. Studies of media representation by gender are rich and varied, often exploring representation across modalities, genres, places, and time spans~\cite{collins_content_2011,smith_inclusion_2024,lauzen_women_nodate,noauthor_gdi_nodate,Haris2023GenderBias}. Many such studies  consist of comparing the proportions of women versus men in media or offering critical analysis of explicit and implicit stereotyping of female characters~\cite{collins_content_2011}. Since the advent of cinema as an art form (including Television), the number of women characters has increased, albeit not always linearly, with setbacks and reversals. Yet in 2024, for the first time, the proportion of films with women as the leading actor exceeded the proportion of women in the US census~\cite{smith_inclusion_2024}.  This phenomenon was not mirrored in broadcast and streaming television, with only 45 percent of shows featuring a female protagonist in 
2023--2024~\cite{lauzen_women_nodate}. 
These patterns persist in comparisons across genre, albeit with different magnitudes. In a study of all 2023 family movies rated G, PG, and PG-13 with a budget of over \$10 million US dollars (82 movies total), 35 percent of leads were female, representing a decrease from 48 percent of all such films released in 2019~\cite{noauthor_gdi_nodate}. Similarly, from 2009--2018, in the 211 major science fiction films with wide releases (1,000+ theaters) only 29 (14 percent) had a female solo lead and 66 (31 percent) had a female co-lead \cite{bbc_wmc_superpowering_2018}. 

Compounding the relative lack of major female roles on-screen is the problematic nature of how women are depicted. 
Numerous studies have examined the extent to which prevalent gender representations sexualize and objectify women, promote stereotyped gender norms, and/or endorse sexist beliefs~\cite{santoniccolo_gender_2023,hentschel_multiple_2019,tremmel_gender_2023,smith_power_2019,koburtay_congruity_2019}. Negative portrayals of women do not arise spontaneously, nor can they be attributed solely to their authors: they often reflect negative beliefs embedded in society.\footnote{A 2024 nationwide survey of the United States (US) reported that only 58\% of Americans believe women are treated with dignity and respect in the US, the lowest among similar countries per the OECD index. Disconcertingly, there is a significant gender gap in response to this particular question in the Gallup survey, with 49 percent of women and 68 percent of men believing women are treated with dignity and respect. Across other nations, the portion of people espousing this belief ranges from 21 percent (Turkey, officially the Republic of T\"{u}rkiye) to 93 percent (Norway)~\cite{inc_gender_2025}.} Negative beliefs about women percolate through varied contexts, leading to widespread, yet sometimes subtle, disparities. In leadership studies, for example, men and women often rate female leaders as demonstrating significantly less agency than male leaders~\cite{koburtay_congruity_2019,smith_power_2019}. Furthermore, in a recent study of public opinion polls from 40 countries regarding men and women, the majority of respondents from all countries rated men as more agentic than women, and women as more communal than men~\cite{Nater2026GenderStereotypes}. These observations make examining individual traits across many fictional characters (as in our dataset) a powerful tool for surfacing implicit biases. 

Media representations influence beliefs, norms, and other social phenomena far beyond the realm of gender. Fictional stories which children and young adults identify as personally salient can impact identity formation~\cite{breen_movies_2017}. \citet{dill-shackleford_connecting_2016} propose a ``dual empathy'' pathway wherein viewers process feelings and thoughts alongside fictional characters. Transportation theory extends narrative theory and posits that people who experience greater
narrative transportation in response to fictional and
non-fictional media are more likely to change their
behaviors and beliefs and identify fewer factual
inconsistencies in that media~\cite{breen_movies_2017}. A small meta-analysis of three studies demonstrates that more frequent television viewing by children correlates with expression of more stereotypes~\cite{ward_media_2020}. The genre and quality of the stories further mediates this effect. For example, a 2014 study of 132 mothers demonstrated a correlation between more frequent viewing of superhero movies by their pre-schoolers and male-stereotyped structured play (e.g., fighting and miming weapon usage)~\cite{coyne_its_2014}. 

 Here, gender expression\footnote{We recognize that a binary conception of gender is not a true representation of the diverse spectrum of gender identities~\cite{APA2015,Hyde2019_future_sex_gender_psychology}, and is a limitation of our work. We focus on male and female classification, based on data availability of gender in film. We exclude $2$ non-binary characters due to sample size limitations, and we refer to classification by gender as the character's ``canonical gender'', described in ``Methods''.} 
 may relate to character's outward physical characteristics like speech, mannerisms, manner of dress~\cite{WorkmanJohnson1993}, and style, or their outward hobbies, interests, and activities~\cite{Lee2018}. Gender expression may also reflect how a character relates with other people in their friendships, romantic relationships~\cite{Thomeer2020} or broader roles in society such as their career~\cite{allison2010heroes}.


\section{Methods}

We start with data collected from the Open Psychometrics survey `Statistical ``Which Character'' Personality Quiz' which comprised the measurement of $464$ traits across $2,000$ characters, with over $72$M ratings total~\cite{openpsychometrics}. Table~\ref{tab:archetypometrics.datasets.cleaned} describes this dataset. Each survey tests for each character a sample of differential ratings on semantic pairs like \traitlinksimple{emotional}{logical} and \traitlinksimple{wild}{tame}. In our previous work~\cite{dodds2025archetypometrics}, we identified three primary archetypes or \textbf{essential dimensions} through dimensionality reduction via singular vector decomposition: 
\archetypesemdiff{1},
\archetypesemdiff{2},
and 
\archetypesemdiff{3}.
In order, these primary archetypes describe the most variance in the essential directions in character and trait space, across characters and their associated trait ratings. We also identified three secondary archetypes: 
\archetypesemdiff{4},
\archetypesemdiff{5},
and
\archetypesemdiff{6}.

For the present analysis, we label each character by gender based on each character's canonical gender; that is, the gender of the character reflected in pronoun usage across Wikipedia~\cite{wikipedia}, Fandom~\cite{fandom}, and IMDB~\cite{imdb}. 
We are able to 
categorize 1,999 out of 2,000 characters based on a binary conception of gender. Our labeled dataset includes 782 canonically female characters and 
1,217 canonically male characters.
The~\characterlinksimple{Alien-the-Alien}{Alien} 
(\storylinksimple{Alien}{Alien})
is the sole canonically genderless character.

We then examine the distribution of each primary and secondary archetype by gender. Next, we compare the distribution of each primary and secondary archetype to the \traitlinksimple{masculine}{feminine} trait pair. In our original work, this specific trait ranked $7^{\textrm{th}}$ in importance of $434$ traits in describing variance among characters. We specifically explore the $25^{\textrm{th}}$, $50^{\textrm{th}}$, and $75^{\textrm{th}}$ percentiles of the median \traitlinksimple{masculine}{feminine} trait score across each archetype by gender. We use a moving window of $n=100$ characters to calculate these three percentiles. 
\begin{table*}[t]
\centering
\renewcommand{\arraystretch}{1.2}
\setlength{\tabcolsep}{8pt}
\begin{tabular}{cccc|cc|cc}
\hline
\rowcolor{headerpop}
\multicolumn{4}{c|}{\textbf{Overall numbers}} &
\multicolumn{2}{c|}{\textbf{Ratings per trait}} &
\multicolumn{2}{c}{\textbf{Ratings per character}} 
\\
 \rowcolor{headerpop}
\textbf{Stories} & \textbf{Traits} & \textbf{Characters} & \textbf{Total ratings} &
\textbf{Average} & \textbf{Range} &
\textbf{Average} & \textbf{Range} \\
\hline
341 & 464 & 2000 & 72,099,735 &
155,387.4 & 25,427--458,342 &
36,049.9 & 3,870--300,067 \\
\hline
\end{tabular}
\caption{
Summary of characters, stories, ratings, and traits.
Data from the
\href{https://openpsychometrics.org}{Open Source Psychometrics Project}~\cite{openpsychometrics}.
}
\label{tab:archetypometrics.datasets.cleaned}
\end{table*}

\section{Results}

The distributions of archetypes by gender vary. In Figure~\ref{fig:archetype_histogram}, we demonstrate using the Kolmogorov-Smirnov test that all distributions between male and female characters by primary and secondary archetypes are statistically significantly different ~\cite{massey_jr_kolmogorov-smirnov_1951} with $p<0.01$.

\subsection{Primary archetypes and canonical gender}

In Figure~\ref{fig:archetype_histogram}A, both canonically male and female characters tend towards \archetype{Hero}, with female characters ($2.09$) exceeding male characters ($1.41$) in their median \archetype{Hero} value. In Figure~\ref{fig:archetype_histogram}B, 
canonically male characters trend slightly towards \archetype{Demon} whereas female characters trend towards \archetype{Angel}. 
In Figure~\ref{fig:archetype_histogram}C, 
both canonically male and female characters trend towards \archetype{Adventurer}, with female characters ($0.64$) exceeding male characters ($0.25$) in their median \archetype{Adventurer} value. 

\subsection{Secondary archetypes and canonical gender}

\begin{figure*}[t]
  \centering
  \includegraphics[width=\textwidth]{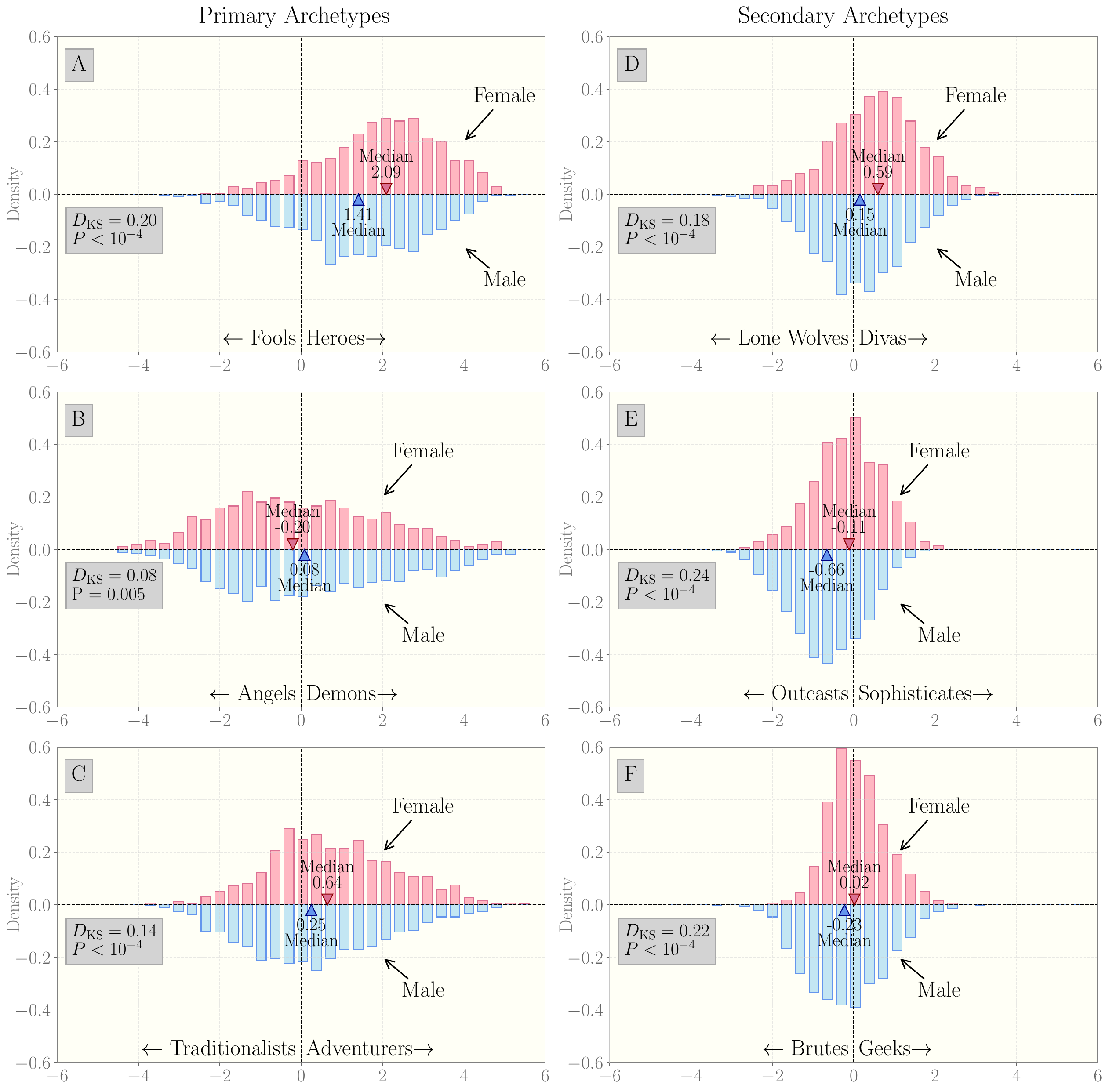} 
  \caption{\textbf{Histograms describing the primary---\protect\archetypesemdiff{1},\protect\archetypesemdiff{2}, and \protect\archetypesemdiff{3}---and secondary---\protect\archetypesemdiff{4},\protect\archetypesemdiff{5}, \protect\archetypesemdiff{6}---archetypes by gender.} A triangle marks the median of each distribution. We also calculate the Kolmogorov-Smirnov~\cite{massey_jr_kolmogorov-smirnov_1951} non-parametric test-statistic demonstrating differences in male versus female distributions; all distributions exhibit significant differences $p<0.01$. The lefthand column of this figure demonstrates primary archetypes and the righthand column demonstrates secondary archetypes.}
  \label{fig:archetype_histogram}
\end{figure*}

In Figure~\ref{fig:archetype_histogram}D, both canonically male and female characters tend towards \archetype{Diva}; however, female characters ($0.59$) exceed male characters ($0.15$) in their median \archetype{Diva} value. 
In Figure~\ref{fig:archetype_histogram}E, both canonically male and female characters tend towards \archetype{Outcast}; however, male characters tend towards \archetype{Outcast} more strongly (as a negatively weighted archetype, the male median score of $-0.66$ exceeds the female median score of $-0.11$). 
In Figure~\ref{fig:archetype_histogram}F, canonically female characters tend slightly towards the \archetype{Geek} archetype ($0.02$), whereas male characters tend towards the \archetype{Brute} archetype ($-0.23$).

\subsection{Archetypes and the \protect\traitlinksimple{masculine}{feminine} trait}

\begin{figure*}[t]

\includegraphics[width=\textwidth]{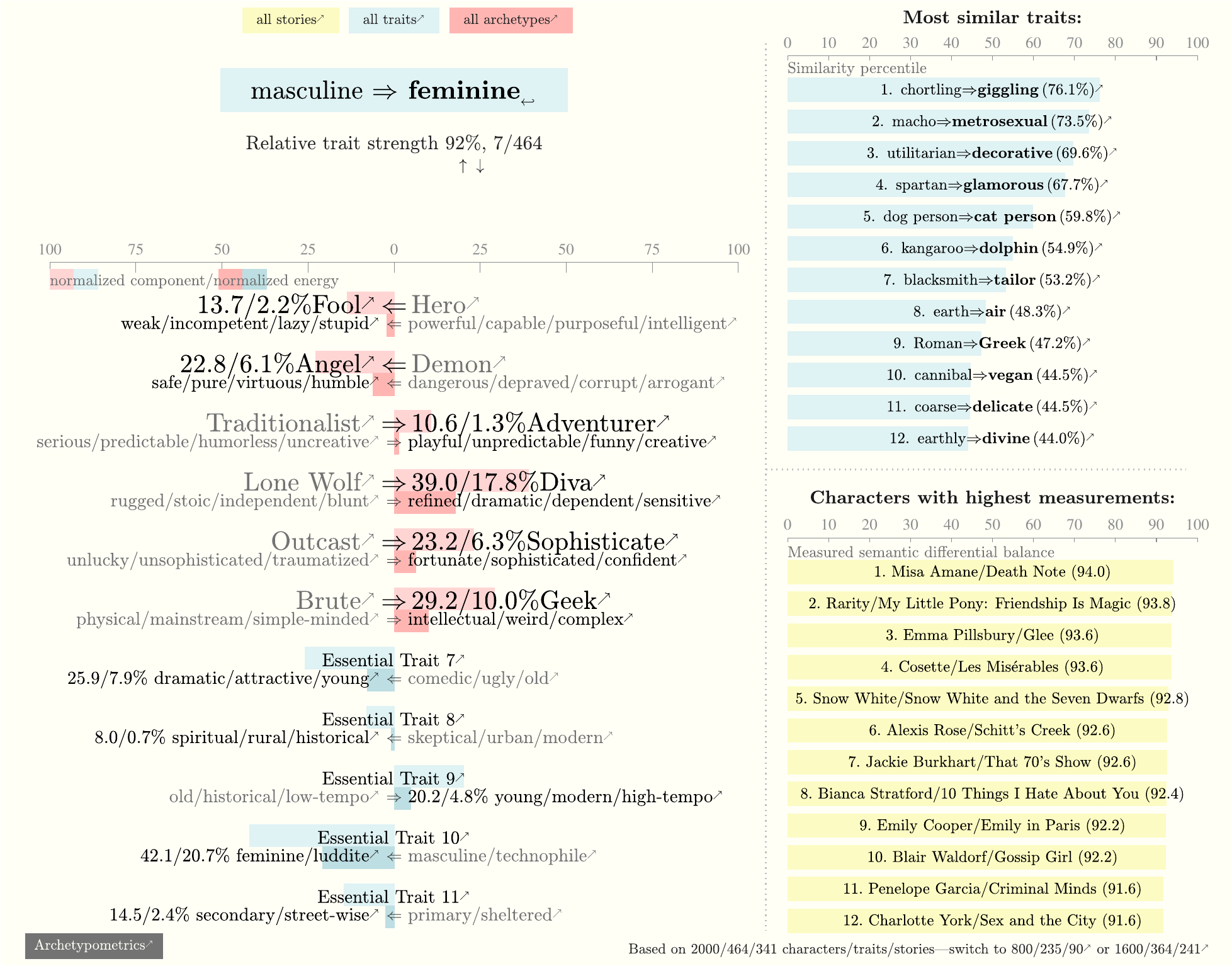}

\caption{
    Card for the \traitlinksimple{masculine}{feminine} trait.
    See Ref.~\cite{dodds2025archetypometrics} for full description.
}
\label{fig:masculine-feminine}
\end{figure*}

\begin{figure*}[t]

\includegraphics[width=\textwidth]{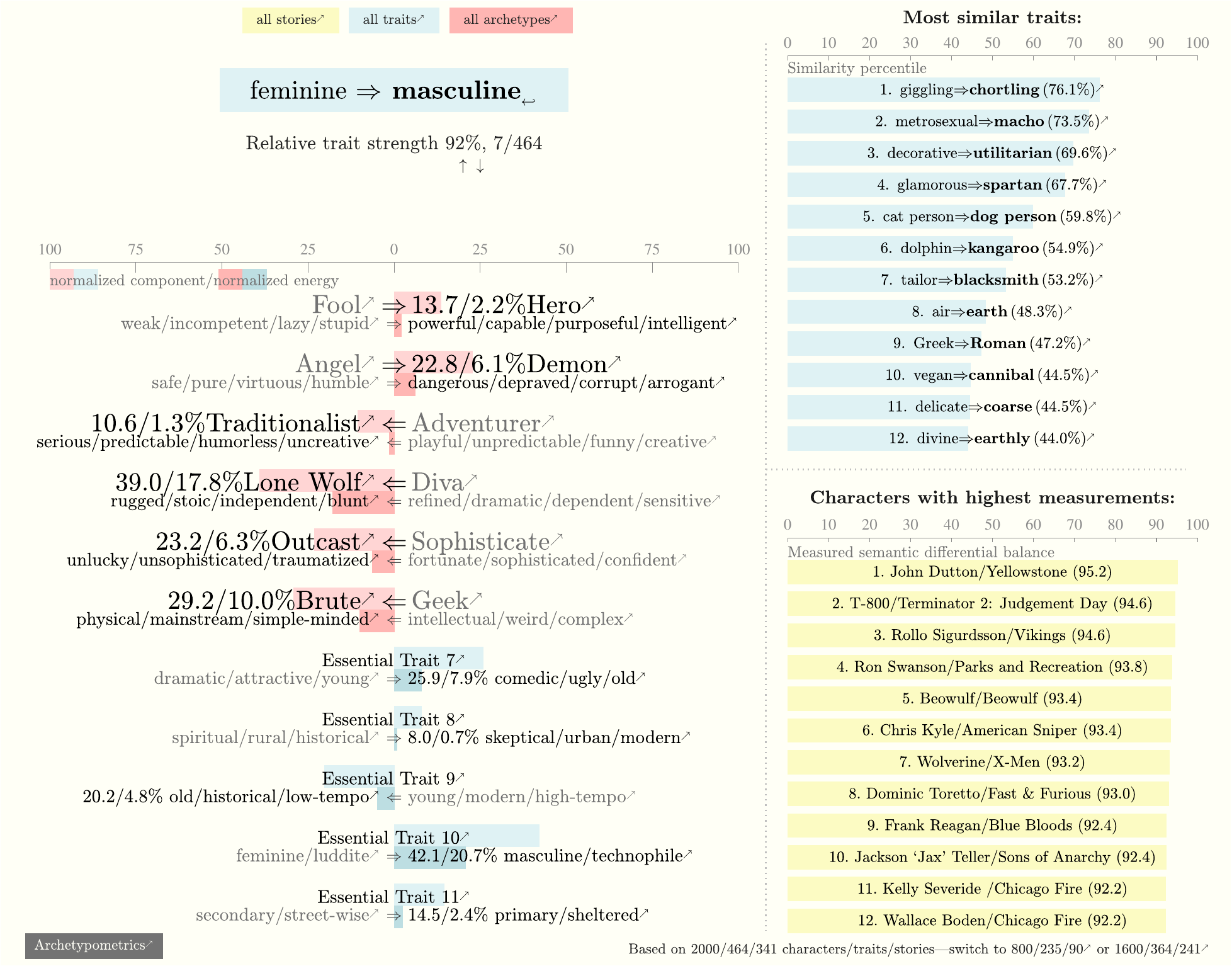}

\caption{
    Card for the \traitlinksimple{feminine}{masculine} trait,
    reverse of the card in~\cite{fig:masculine-feminine}.
    See Ref.~\cite{dodds2025archetypometrics} for full description.
}
\label{fig:feminine-masculine}
\end{figure*}

Across archetype space, the \traitlinksimple{masculine}{feminine} trait ranks as the $7^{\textrm{th}}$  largest trait describing variance in each character's archetype. 
Figures~\ref{fig:masculine-feminine} and~\ref{fig:feminine-masculine} show `trait cards' which display the normalized archetype components of the \traitlinksimple{masculine}{feminine} and \traitlinksimple{feminine}{masculine} traits, respectively. These figures are reflections of one another; colloquially, they depict two sides of the same coin. For example, the masculine trait is most similar to the chortling trait in the \traitlinksimple{giggling}{chortling}, whereas the feminine trait is most similar to the giggling trait in the same pair. 
The masculine trait aligns with \archetype{Hero}, \archetype{Demon}, \archetype{Traditionalist}, \archetype{Lone Wolf}, \archetype{Outcast}, and \archetype{Brute} archetypes. The feminine trait aligns with the \archetype{Fool}, \archetype{Angel}, \archetype{Adventurer}, \archetype{Diva}, \archetype{Sophisticate}, and \archetype{Geek} archetypes. Importantly, trait and archetype alignment within canonical gender groups does not necessarily reflect individual trait scores and archetype alignment (discussed in following section).  

The top three most masculine characters are \characterlinksimple{Yellowstone-John-Dutton}{John Dutton} from the Western television series \storylinksimple{Yellowstone}{Yellowstone}, \characterlinksimple{Terminator-2-Judgement-Day-T-800}{T-800} from the sci-fi/horror movie \storylinksimple{Terminator-2-Judgement-Day}{Terminator 2: Judgement Day} and \characterlinksimple{Vikings-Rollo-Sigurdsson}{Rollo Sigurdsson} from the historical drama \storylinksimple{Vikings}{Vikings}. The top three most feminine characters are \characterlinksimple{Death-Note-Misa-Amane}{Misa Amane} from the anime series \storylinksimple{Death Note}{Death Note}, \characterlinksimple{My-Little-Pony-Friendship-Is-Magic-Rarity}{Rarity} from \storylinksimple{My-Little-Pony-Friendship-Is-Magic}{My Little Pony: Friendship Is Magic} animate series, and \characterlinksimple{Glee-Emma-Pillsbury}{Emma Pillsbury} from the comedy drama \storylinksimple{Glee}{Glee}. Interestingly, a character can rate most highly among all characters on a trait without that specific trait comprising the largest portion of that individual character; for example, \characterlinksimple{Yellowstone-John-Dutton}{John Dutton} rates highest on the \traitlinksimple{feminine}{masculine} trait in the direction of masculinity, but the trait on which survey respondents rank him at the top is \traitlinksimpleright{queer}{straight}.


\begin{figure}[t]

\includegraphics[width=\columnwidth]{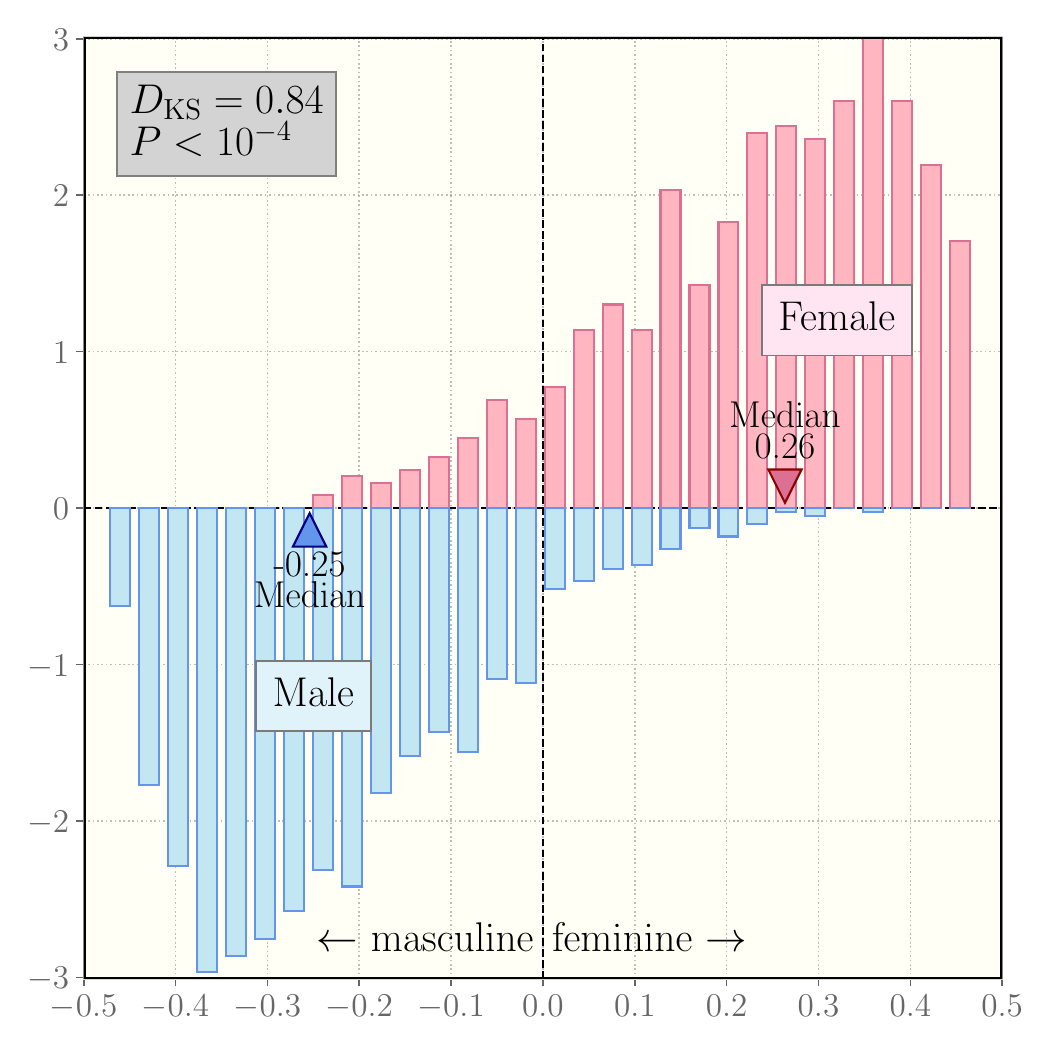}

\caption{
    \textbf{Distributions of canonically male and female characters
    along the 
    \traitlinksimple{masculine}{feminine}
    trait.} This histogram shows the spread of ratings of the singular \traitlinksimple{masculine}{feminine} for each character in the dataset. The non-parametric Kolgomorov-Smirnov test strongly indicates a difference in these gender distributions. Canonically female characters tend to rate much more highly in the direction of feminine; likewise, canonically male characters tend to rate much more highly in the direction of masculine.  
}
\label{fig:male-female-masculine-feminine-distributions}
\end{figure}

\subsection{Primary archetypes, canonical gender, and the \protect\traitlinksimple{masculine}{feminine} trait}

Next, we explore the connections between archetypes and the \traitlinksimple{masculine}{feminine} trait. 
We first show how ratings of canonically male and female characters are distributed for the \traitlinksimple{masculine}{feminine} trait, 
shown in Figure~\ref{fig:male-female-masculine-feminine-distributions}. Not surprisingly, the median trait score for canonically female characters falls firmly within the feminine direction, whereas the median for canonically male characters is in the masculine direction. 

\begin{figure*}[t]
  \centering
  \includegraphics[width=\textwidth]{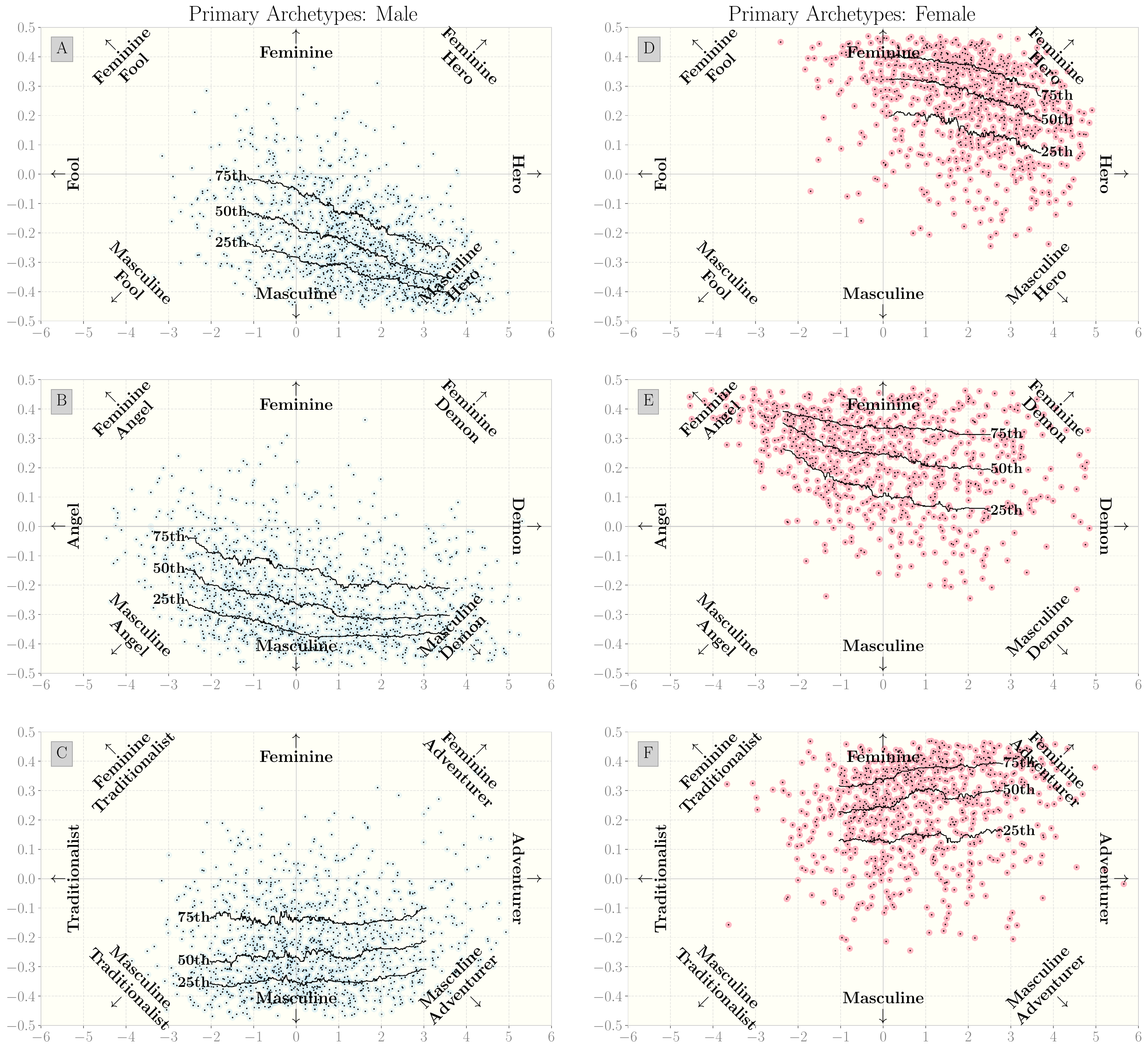} 
  \caption{\textbf{The \traitlinksimple{masculine}{feminine} trait pair versus the primary archetypes by gender.} The panel on the left of this figure depicts male characters, their primary archetype ratings, and their ratings for the \traitlinksimple{masculine}{feminine} trait. The panel on the right of this figure depicts female characters, their primary archetype ratings, and their ratings for the \traitlinksimple{masculine}{feminine} trait. The percentile lines depict the $25^{\textrm{th}}$, $50^{\textrm{th}}$, and $75^{\textrm{th}}$ percentiles for the rolling median \traitlinksimple{masculine}{feminine} trait scores in $n=200$ windows.}
  \label{fig:scatterplot_primary}
\end{figure*}

In Figure~\ref{fig:scatterplot_primary}A, we observe a positive relationship between the \archetypesemdiff{1} differential and the masculine trait scores for canonically male characters, with exemplar characters like \characterlinksimple{Star-Trek-The-Next-Generation-Jean-Luc-Picard}{Jean-Luc Picard} from the sci-fi television series \storylinksimple{Star-Trek-The-Next-Generation}{Star Trek: The Next Generation} and \characterlinksimple{Battlestar-Galactica-William-Adama}{William Adama} from \storylinksimple{Battlestar-Galactica}{Battlestar Galactica}.  

In Figure~\ref{fig:scatterplot_primary}B, we similarly observe a positive relationship between the \archetypesemdiff{2} differential and the masculine trait scores for canonically male characters, for example, \storylinksimple{Toy-Story}{Toy Story's} \characterlinksimple{Toy-Story-Sid-Phillips}{Sid Phillips}. 

In Figure ~\ref{fig:scatterplot_primary}C, we observe an equivocal relationship between the \archetypesemdiff{3} archetype pair and the \traitlinksimple{masculine}{feminine} trait. 

For canonically female characters, we observe many of the same patterns in Figure~\ref{fig:scatterplot_primary}D--F. In Figure~\ref{fig:scatterplot_primary}D, we note a positive relationship between the
\archetypesemdiff{1}
archetype pair and the masculine trait scores for canonically female characters, with female heroes like \characterlinksimple{A-Series-of-Unfortunate-Events-Violet-Baudelaire}{Violet Baudelaire} from \storylinksimple{A-Series-of-Unfortunate-Events}{A Series of Unfortunate Events} and \characterlinksimple{Harry-Potter-Hermione-Granger}{Hermione Granger} from \storylinksimple{Harry Potter}{Harry Potter}. 

Similarly, 
in Figure~\ref{fig:scatterplot_primary}E, 
we observe a positive relationship between the 
\archetypesemdiff{2} pair
and the masculine trait scores for canonically female characters, like \storylinksimple{Mean-Girls}{Mean Girls'} Queen Bee bully \characterlinksimple{Mean-Girls-Regina-George}{Regina George}. 

In Figure~\ref{fig:scatterplot_primary}F, we observe a slightly different relationship between canonically male and female characters and their \traitlinksimple{masculine}{feminine} trait scores for the \archetypesemdiff{3} pair; canonically female characters trend slightly more feminine in the direction of the \archetype{Adventurer} archetype (like social media guru and entrepreneur \characterlinksimple{Ted-Lasso-Keeley-Jones}{Keeley Jones} in \storylinksimple{Ted Lasso}{Ted Lasso}).

For the two leading primary archetypes, both male and female characters who skew towards the \archetype{Hero} archetype or the \archetype{Demon} archetype also rate more highly on the masculine trait. For the third primary archetype pair, \archetypesemdiff{3}, there is a slight positive relationship between femininity and the \archetype{Adventurer} archetype for male characters, and a stronger positive relationship between femininity and the \archetype{Adventurer} archetype for female characters. 

Our supplement contain further primary archetype ousioograms by canonical gender in Supplementary Figures \ref{fig:ousiogram-feminine-masculine-male-01}, ~\ref{fig:ousiogram-feminine-masculine-male-02}, ~\ref{fig:ousiogram-feminine-masculine-male-03},  ~\ref{fig:ousiogram-feminine-masculine-female-01}, ~\ref{fig:ousiogram-feminine-masculine-female-02}, and ~\ref{fig:ousiogram-feminine-masculine-female-03}, with each distribution's convex hull annotated with character names and series. 

\begin{figure*}[t]
  \centering
  \includegraphics[width=\textwidth]{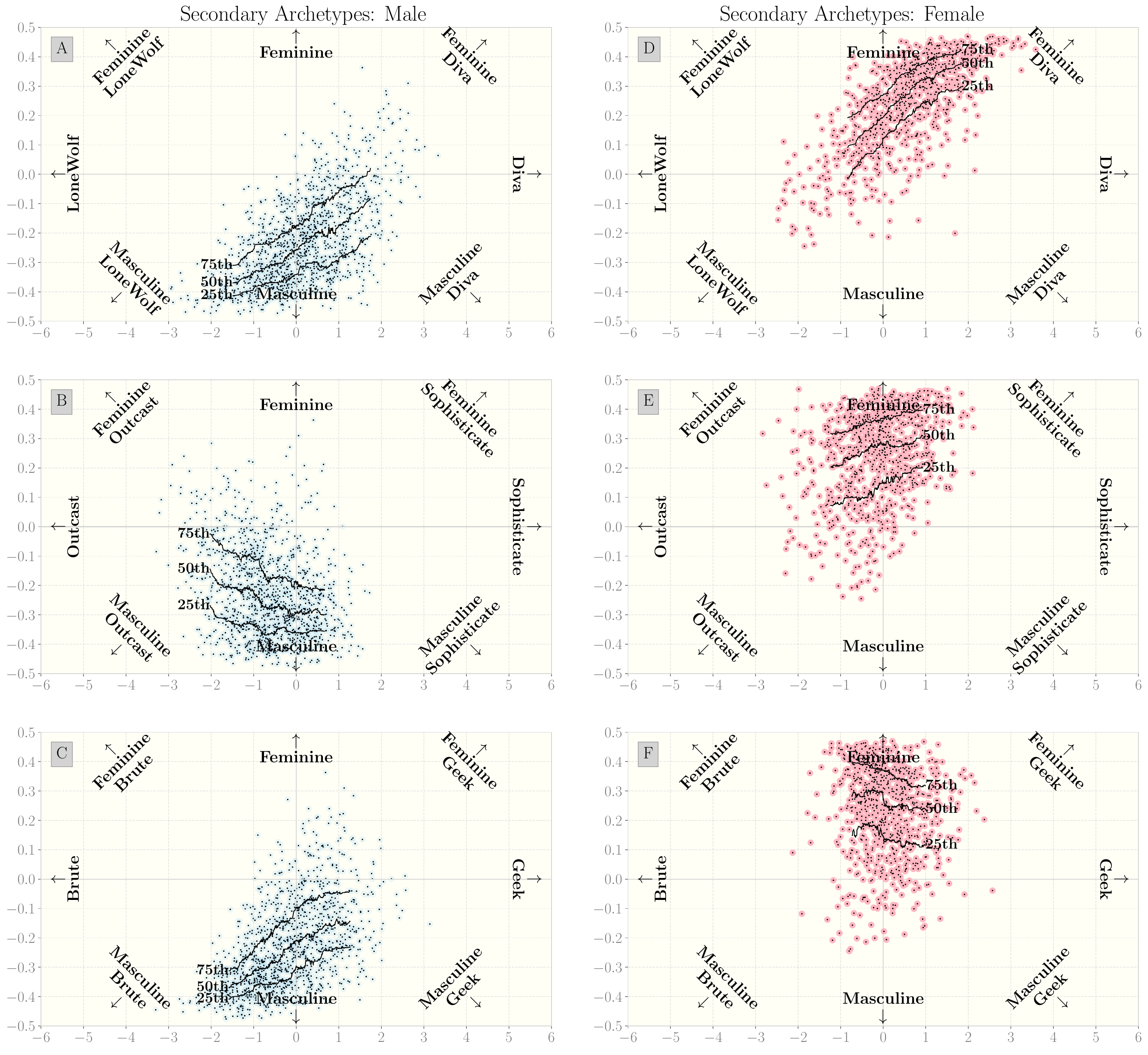} 
  \caption{\textbf{The \traitlinksimple{masculine}{feminine} trait pair versus the secondary archetypes by gender.} The panel on the left of this figure depicts male characters, their secondary archetype ratings, and their ratings for the \traitlinksimple{masculine}{feminine} trait. The panel on the right of this figure depicts female characters, their secondary archetype ratings, and their ratings for the \traitlinksimple{masculine}{feminine} trait. The percentile lines depict the $25^{\textrm{th}}$, $50^{\textrm{th}}$, and $75^{\textrm{th}}$ percentiles for the rolling median of \traitlinksimple{masculine}{feminine} trait scores in the $n=200$ windows.}
  \label{fig:scatterplot_secondary}
\end{figure*}

\subsection{Secondary archetypes, canonical gender, and the \protect\traitlinksimple{masculine}{feminine} trait}

In Figure~\ref{fig:scatterplot_secondary}A, we observe a positive relationship between the 
\archetypesemdiff{4} pair
\traitlinksimple{masculine}{feminine}
trait scores for canonically male characters, with characters like \storylinksimple{Glee}{Glee's} \characterlinksimple{Glee-Kurt-Hummel}{Kurt Hummel} and \storylinksimple{Unbreakable-Kimmy-Schmidt}{Unbreakable Kimmy Schmidt's} \characterlinksimple{Unbreakable-Kimmy-Schmidt-Titus-Andromedon}{Titus Andromedon} displaying a relatively high femininity score alongside the \archetype{Diva} archetype.  

In Figure~\ref{fig:scatterplot_secondary}B, we observe a positive relationship between the \archetype{Sophisticate} and the masculine trait scores for canonically male characters, exemplified by quintessential comedic \archetype{Outcast} characters \characterlinksimple{Veep-Gary-Walsh}{Gary Walsh} in \storylinksimple{Veep}{Veep} and \characterlinksimple{Arrested-Development-Buster-Bluth}{Buster Bluth} in \storylinksimple{Arrested-Development}{Arrested Development} who have relatively high femininity scores (and are -- interestingly -- played by the same actor). 

In Figure~\ref{fig:scatterplot_secondary}C, we observe a positive relationship between the \archetype{Brute} archetype and the masculine trait score for canonically male characters, as in the anti-hero, conflicted \characterlinksimple{Mr-Robot-Elliot-Alderson}{Elliot Alderson} from \storylinksimple{Mr-Robot}{Mr. Robot}. 

For canonically female characters, we observe somewhat different patterns in Figure~\ref{fig:scatterplot_secondary}D--F. In Figure~\ref{fig:scatterplot_secondary}D, we note a similar positive relationship between the \archetype{Diva} archetype and the feminine trait scores for canonically female characters, for example \characterlinksimple{Glee-Rachel-Berry}{Rachel Berry} from \storylinksimple{Glee}{Glee} and \characterlinksimple{That-70s-Show-Jackie-Burkhart}{Jackie Burkhart} from \storylinksimple{That-70s-Show}{That 70's Show}. Unlike male characters, we observe a positive relationship between \archetype{Sophisticate} and feminine trait scores for canonically female characters; a well-known example is \characterlinksimple{Parks-and-Recreation-Donna-Meagle}{Donna Meagle} from the comedy television series \storylinksimple{Parks-and-Recreation}{Parks and Recreation}. Conversely, a well-known outcast who tends to be more masculine than other canonically female characters is Mrs. Maisel's manager \characterlinksimple{The-Marvelous-Mrs-Maisel-Susie-Myerson}{Susie Myereson} from \storylinksimple{The-Marvelous-Mrs-Maisel}{The Marvelous Mrs. Maisel}. 

In Figure~\ref{fig:scatterplot_secondary}F, we observe a slight positive relationship between the \archetype{Geek} and masculine trait scores for canonically female characters; for example, \characterlinksimple{The-Breakfast-Club-Allison-Reynolds}{Allison Reynolds} from \storylinksimple{The-Breakfast-Club}{The Breakfast Club}.

In summary, for the secondary archetypes, we observe more extreme relationships than among primary archetypes. For the\archetypesemdiff{4} archetype pairing, we notice a strong relationship moving from \archetype{Lone Wolf} to \archetype{Diva} archetypes and the feminine trait. For both\archetypesemdiff{5} and\archetypesemdiff{6} archetype pairs, we observe inverse relationships between the \traitlinksimple{masculine}{feminine} trait and character gender. For the\archetypesemdiff{5} moving from \archetype{Outcast} to \archetype{Sophisticate} archetype yields higher masculinity ratings for male characters, whereas female characters yield higher femininity ratings. Conversely, for the\archetypesemdiff{6} archetype, moving from \archetype{Brute} to \archetype{Geek} yields higher femininity ratings for male characters, whereas female characters yield higher masculinity ratings. 

Our supplement contain further secondary archetype ousioograms by canonical gender in Supplementary Figures \ref{fig:ousiogram-feminine-masculine-male-04}, ~\ref{fig:ousiogram-feminine-masculine-male-05}, ~\ref{fig:ousiogram-feminine-masculine-male-06},  ~\ref{fig:ousiogram-feminine-masculine-female-04}, ~\ref{fig:ousiogram-feminine-masculine-female-05}, and ~\ref{fig:ousiogram-feminine-masculine-female-06}, with each distribution's convex hull annotated with character names and series.

\section{Discussion}

While the discovered archetypal patterns initially seem to refute traditional gender representations, closer examination of the \traitlinksimple{masculine}{feminine} character trait reveals a persistent adherence to conventional gender expectations. Importantly, our dataset is modern, Western (and often Anglophone) popular film and television, so we consider the following discussion less generalizable outside of this specific cultural realm. In our conclusion, we propose future works that address these limitations.

In previous work, we discover a bias in the entire dataset across both canonical genders ~\cite{dodds2025archetypometrics} toward characters that demonstrate competence, self-control, and drive, even though not all the works feature heroic characters.\footnote{Psychologists have widely studied the phenomena of hero bias in both surveys and psychological experiments. Refer to Scott and Goethals' 2010 book ``Heroes: What They Do and Why We Need Them''~\cite{allison2010heroes}.} 
For example, there are no \archetype{Hero} archetypes~\cite{dodds2025archetypometrics} among characters in the long-running dark comedy,
\storylinksimple{Its-Always-Sunny-in-Philadelphia}{It's Always Sunny in Philadelphia}~\cite{wikipedia_sunny}. 
As evidenced in Figure~\ref{fig:archetype_histogram}, the hero bias is apparent in both canonical genders. Consequently, examining archetypes \textit{by} canonical gender uncovers differences in the extent to which male and female characters adhere to archetypes. 

\begin{figure}[h]
  \centering
  \includegraphics[width=0.47\textwidth]{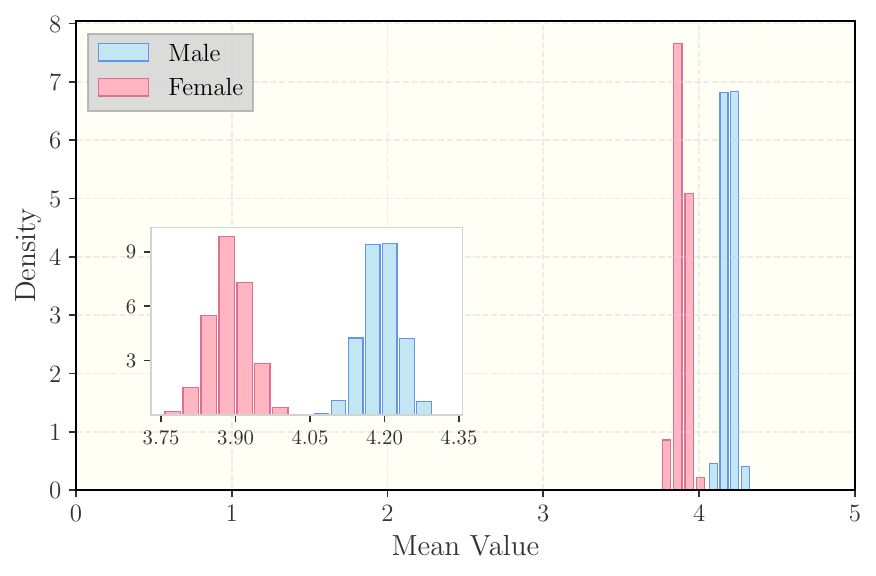} 
  \caption{
  \textbf{Bootstrapped Radius of Gyration for Male and Female Characters, $n$ = 500 samples each gender, averaged over $n$ = 5,000 iterations.} Bootstrapped sampling of radius of gyration to balance the uneven number of characters by gender; female characters exhibit significantly lower variance than male characters across trait ratings.}
  \label{fig:bootstrap_radius}
\end{figure}

Globally, our results indicate a prevalent canonical gender bias reflected by significantly different histograms of primary and secondary archetypes by gender in Figure~\ref{fig:archetype_histogram}. 
The variance of trait ratings of female and male characters, using bootstrapped sampling methods to balance the uneven number of characters by gender, reveal significantly less variance among female characters than male (Figure~\ref{fig:bootstrap_radius}). This observation reflects the presence of gender biases and stereotypes noted in sociological and psychological studies~\cite{santoniccolo_gender_2023,hentschel_multiple_2019,tremmel_gender_2023,smith_power_2019,koburtay_congruity_2019}, but does not provide insight into the quality or nature of these biases. 

At the level of individual archetype distributions, the confirmation or refutation of gender stereotypes might highlight the cultural importance of any particular stereotype. Alternatively, it may simply reflect the relative ease with which the \traitlinksimple{masculine}{feminine} trait can be assigned to certain characters. Ellemers' review of the prevalence and persistence of gender stereotypes indicates that male stereotypes tend towards task completion and performance, while female stereotypes tend to focus on social relationships~\cite{Ellemers2018}. 

Firstly, the range of \traitlinksimple{masculine}{feminine} ratings for canonically male characters is wider than for canonically female characters, suggesting stricter personality stereotyping of female characters with male characters appearing to have greater latitude in trait expression. Notably, canonically female characters score higher on heroism and adventurousness than their male counterparts, defying traditional representations of women as unworldly, incapable ``damsels in distress''~\cite{maity_2014,putri_breaking_2024}.  Yet regardless of canonical gender, more heroic characters consistently rate higher on masculinity; in the case of adventuresome characters, the relationship between masculinity and femininity is ambivalent. It is possible that these divergences reflect a societal bias towards representing women as weak or less agentic~\cite{phelan_prejudice_2010} \textit{or} it could reflect that the conception of hero since classical antiquity has centered on masculinity~\cite{FoxhallWhenMenWere2013} and  --- especially within an oral tradition --- reflect the cultural traditions of society~\cite{nagy2013epichero}. The stereotype of ``damsel in distress'' surfaces again in the trait-level trends for the \archetypesemdiff{2} archetype with angelic characters of both canonical genders maintaining higher femininity ratings (thus, conforming to traditional stereotypes); for example \characterlinksimple{Pride-and-Prejudice-Jane-Bennet}{Jane Bennet} from \storylinksimple{Pride-and-Prejudice}{Pride and Prejudice} and \characterlinksimple{30-Rock-Kenneth-Parcell}{Kenneth Parcell} from \storylinksimple{30-Rock}{30 Rock} both display relatively higher femininity ratings compared to other characters of their canonical genders. 

For secondary archetypes, the trends in the \traitlinksimple{masculine}{feminine} trait and traditional gender stereotypes are more apparent. This may reflect the inherently gendered nature of these archetypes in that the \traitlinksimple{masculine}{feminine} trait is ``baked in'' to these characterizations, even if the trait does not predominantly describe that archetype. Among both canonically male and female characters, the \archetype{Diva} archetype trends strongly with femininity. This is true within the \traitlinksimple{masculine}{feminine} trait, but also in their overall portrayal. For example, the top three characters by component size for the archetype \archetype{Diva} are \characterlinksimple{Glee-Rachel-Berry}{Rachel Berry} from \storylinksimple{Glee}{Glee}, \characterlinksimple{New-Girl-Schmidt}{Schmidt} from \storylinksimple{New-Girl}{New Girl} and \characterlinksimple{Good-Place-Tahani-Al-Jamil}{Tahani Al-Jamil} from \storylinksimple{The-Good-Place}{The Good Place}; these characters display extremely stereotypical feminine behavior in their ``prissy'' affect and delivery (a female gender stereotype observed from children's books~\cite{taylor_content_2003} to the workplace~\cite{hu_communicated_2024}). This observation reflects the classical conception of the diva where, in the Italian opera~\footnote{In the cultural lineage of opera, \textit{Castrati} or castrated male opera singers, renowned for their excellent vocal range, predated divas. Though extremely popular in the 16th and 17th centuries, \textit{castrati} were often viewed as problematic and ``feminizing'' by Italian society~\cite{Feldman2008Denaturing} regardless of their artistic excellence. Thus suggesting a longer cultural history of devaluing feminine artists.} of the early 19th century, female opera roles played by the emerging class of divas were, as Rutherford notes, ``...defined by, and designed to support patriarchal hegemony''~\cite{Rutherford2016}.

Among the\archetypesemdiff{5} archetype pairing, the masculine-feminine trend changes for canonically male or female characters. For male characters, \archetype{Outcast} trends with femininity (like \characterlinksimple{Arrested-Development-Buster-Bluth}{Buster Bluth} in \storylinksimple{Arrested-Development}{Arrested Development}, previously mentioned), but for female characters, \archetype{Sophisticate} trends with femininity. The observation among male characters reflects cultural findings on the relative dislike of ``feminine men'' and ``masculine women'' versus individuals who more readily fit their gendered grouping~\cite{leaper_use_1995} and historical depictions in medieval Icelandic literature of labeling men ``unmanly'' leading to blood feuds and defamation lawsuits~\cite{meulengracht1983unmanly}; among canonically female characters, the trends towards femininity and \archetype{Sophisticate} reflect the feminine ideal of austere cleanliness. 

Lastly, we interpret the\archetypesemdiff{6} archetype pairing. For canonically male characters, the well-established relationship with masculinity and physicality~\cite{leone_hegemonic_2018} is on display; men who are more masculine trend strongly towards the Brute archetype. However, the trend towards masculinity and the Geek archetype among canonically female characters appears to depend more on what female characters also lack, similar to Lacan's psychoanalytical theory that womanhood exists as a response to women's lack of masculine biological features (that is, the concept of women is a foil to the concept of man, and not an independent symbolic category~\cite{lacan1958phallus}). The Geek archetype depends on relatively higher ratings among traits like 
\traitlinksimpleright{lumberjack}{mad-scientist},
  \traitlinksimpleright{jock}{nerd},
  and \traitlinksimpleright{physical}{intellectual}~\cite{dodds2025archetypometrics}; 
it is possible that canonically female characters who score lower on these traits also score higher on femininity.
Given the ample scholarship establishing a strong connection between gender representations and societal beliefs~\cite{dill-shackleford_connecting_2016,breen_movies_2017}, the global and local differences in archetypes and the investigation of the \traitlinksimple{masculine}{feminine} trait by gender provide a powerful lens through which to explore both expressions of femininity and prototypical male and female characterizations across visual media. For the \archetypesemdiff{1} archetype that describes the greatest variance in this dataset, heroes of both canonical genders trend towards masculine traits, mirroring established stereotypes documented in social media research\footnote{Memes, while predating the internet, have become an integral part of ``netizen vernacular'' and communicate different messages via their content, form and stance~\cite{shifman2013memes}. Formats like ``Get You A Girl Who Can Do Both'' present a dual image of a woman in a stereotypically feminine or sexually desirable outfit, alongside a male-dominated, machismo profession or hobby~\cite{mason2022get}. While at first this meme format seems empowering, internet users often employ it in a subversive, slightly mocking way~\cite{mason2022get}. 
Similarly, terms like ``girlboss'' evoke a powerful characterization of an entrepreneurial woman, but often center on ``masculine entrepreneurial heroism''~\cite{byrne_giuliani_2025_girlboss}; ``girlboss'', that is, is empowering at first glance, but in fact purports the same sort of limiting and monolithic female leadership made popular in former Facebook CEO Sheryl Sandberg's \textit{Lean In}~\cite{fradley2022so}. Likewise, the popular edict ``she's not like other girls'' focuses on women's implicit focus or explicit desire to distance themselves from other women to gain legitimacy or traction in male dominated spaces, like internet gaming culture~\cite{ruotsalainen_merilainen_2025_girlgamers,Nakandala2017GenderedConversation}} as well as seminal work in film and gender studies. In her book ``Men, Women, and Chain Saws: Gender in the Modern Horror Film'', Carol Clover refers to ``the final girl''---the last surviving female character who fights her way out of the slasher's traps---in 1970s and 1980s slasher/horror films as a \textit{victim}-hero, not simply a hero~\cite{clover2015men}. This characterization centers on her posit that ``female heroes, when they do appear, are masculine in dress and behavior (and often even name)''~\cite{clover2015men}; thus, effeminate female main characters often present as victims, even if they also possess physical courage and other classic hero tendencies. Our findings thus extend Clover's work in a specific, time-dependent context (20th century horror) to a larger grouping of genres and modalities that have persisted well into the 21st century.  

\section{Limitations and Future Work}

Importantly, these stories are film and television shows that are largely produced by western production companies and in English; consequently, any conclusions regarding the relationship between archetype representation and gender representation, norms, and stereotyping apply to this specific, situated media landscape. 

Second, our data is based on survey data where the subjects are self-selecting internet users. Methodologically, internet surveys collect responses from a potentially unrepresentative sample of people, and lack the usual controls of classic surveys~\cite{evans_value_2005,fan_factors_2010,fricker_advantages_2002}. Future work should derive these representations not just from surveys but also from the works of art themselves, and should expand such work to different time periods, mediums, populations, and cultures.  

Additionally, we recognize that our examination of the binary \traitlinksimple{masculine}{feminine} trait can reduce consideration of individual character gender beyond canonical gender; for example, simply because someone is perceived as masculine does not mean that they are devoid of femininity. As representation of transgender and non-binary individuals has increased in media~\cite{mcinroy_transgender_2015}, future work can and should examine works including non-binary characters. 

Finally, we note some methodological advantages for use in future work. The process of grouping characters by trait strength can surface surprising connections. Note that because we can group characters according to traits or archetypes, we can systematically sort the characters of our dataset. This clustering can inspire additional analyses comparing other traits and their relationship with canonical gender, as well as features such as genre, release date, and format (situational comedy versus showcase drama). Furthermore, we can examine the associations between different traits to surface implicit biases, such as the previously mentioned similarity between the \traitlinksimple{giggling}{chortling} trait and masculinity.

\section{Concluding remarks}

The representation of characters in fictional works has wide-ranging social, psychological, and economic impact~\cite{santoniccolo_gender_2023,hentschel_multiple_2019,tremmel_gender_2023,smith_power_2019,koburtay_congruity_2019,jenkins_clean_2018}. This is the first study to systematically consider the representation of predominant archetypes in television and film through the lens of canonical character gender. 
We show that canonically male and female characters differ significantly in their archetypal representations. 

Our novel archetypometric methodology (introduced in \citet{dodds2025archetypometrics}, contextualized in \citet{zimmerman_locality_nodate}, and applied to the exploration of a specific societal construct for the first time here) allows the quantitative exploration of biases and beliefs. We explore structure through the archetypes and the positions of the characters in character space according to their ratings along both the archetypal and original trait dimensions. Furthermore, we explore secondary structure through the associations between these attributes, as we demonstrate here, revealing patterns and implicit biases that are not obvious from viewing the traits or characters individually. In effect, we develop a deep quantitative framework that supports multiple analytical levels, both primary and derivative.

Although primary archetypal dimensions initially appear to challenge some aspects of conventional gender representations namely in the greater heroism and adventurousness of canonically female characters, closer analysis reveals that our results largely adhere to familiar gender norms. Canonically female characters cluster in the feminine direction and canonically male characters in the masculine direction. Across all primary and secondary archetypes, male and female character distributions differ significantly: There is no archetype for which gender is irrelevant.

Normative gender stereotypes persist throughout this database, despite the increase in relative numbers of female characters in leading actor roles~\cite{noauthor_gdi_nodate}. That gender stereotyping has remained largely stable over time suggests that more representation does not entail fewer stereotypical representations, a pattern echoed in persistent gender earnings gaps worldwide, where men continue to out-earn women despite women's increasing labor market share and educational attainment~\cite{UNWomenAfrica_WhyWomenEarnLess2023,EuropeanCommission2024GenderPayGap,RotmanMandel2022_GenderWageGap}, as well as in the continued exclusion of women from the highest levels of corporate leadership (9\% of Fortune 500 CEOs in 2024)~\cite{Catalyst_WomenCEOs_2026}. Our findings suggest that archetypes encode multiple, distinct mechanisms through which gender norms are reproduced across fictional and social domains. Changing who appears on screen, it turns out, does not change what they are expected to be.


%% file: inputs/localized/tabessential_archetypes200_basevarexpl_N2000_dim01.tex
24.4

%% file: inputs/localized/tabessential_archetypes200_basevarexpl_N2000_dim02.tex
20.4

%% file: inputs/localized/tabessential_archetypes200_basevarexpl_N2000_dim03.tex
14.6

%% file: inputs/localized/tabessential_archetypes200_basevarexpl_N2000_dim04.tex
6.2

%% file: inputs/localized/tabessential_archetypes200_basevarexpl_N2000_dim05.tex
5.1

%% file: inputs/localized/tabessential_archetypes200_basevarexpl_N2000_dim06.tex
3.8

%% file: archetypes-and-gender.acknowledgments.tex
The authors are grateful for
National Science Foundation Award \#2242829
(Science of Online Corpora, Knowledge, and Stories),
foundational support from MassMutual,
and
an anonymous philanthropic gift.

%% file: archetypes-and-gender.biblio.tex
\bibliography{\filenamebase}

%% file: archetypes-and-gender.supplementary.tex
\begin{figure*}[t]

\includegraphics[width=\textwidth]{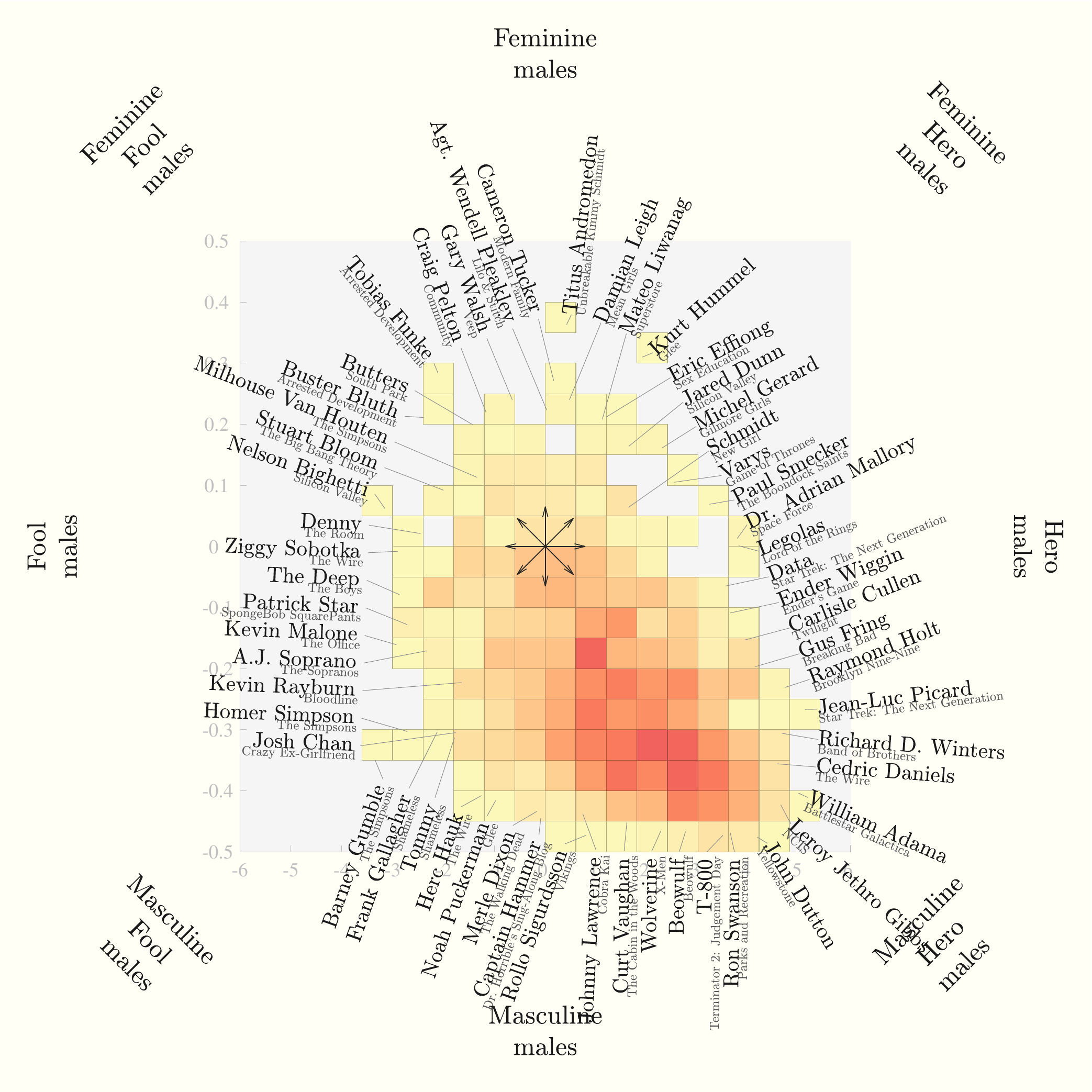}

\caption{
This ousiogram depicts canonically male characters with their \protect\traitlinksimple{masculine}{feminine} score on the vertical axis and their \protect\archetypesemdiff{1} rating on the horizontal axis. The color of each block corresponds to the number of characters that demonstrates scores within the block's range; darker colors indicate more characters fall within that range than lighter color blocks, which may only contain one character.}
\label{fig:ousiogram-feminine-masculine-male-01}
\end{figure*}

\begin{figure*}[t]

\includegraphics[width=\textwidth]{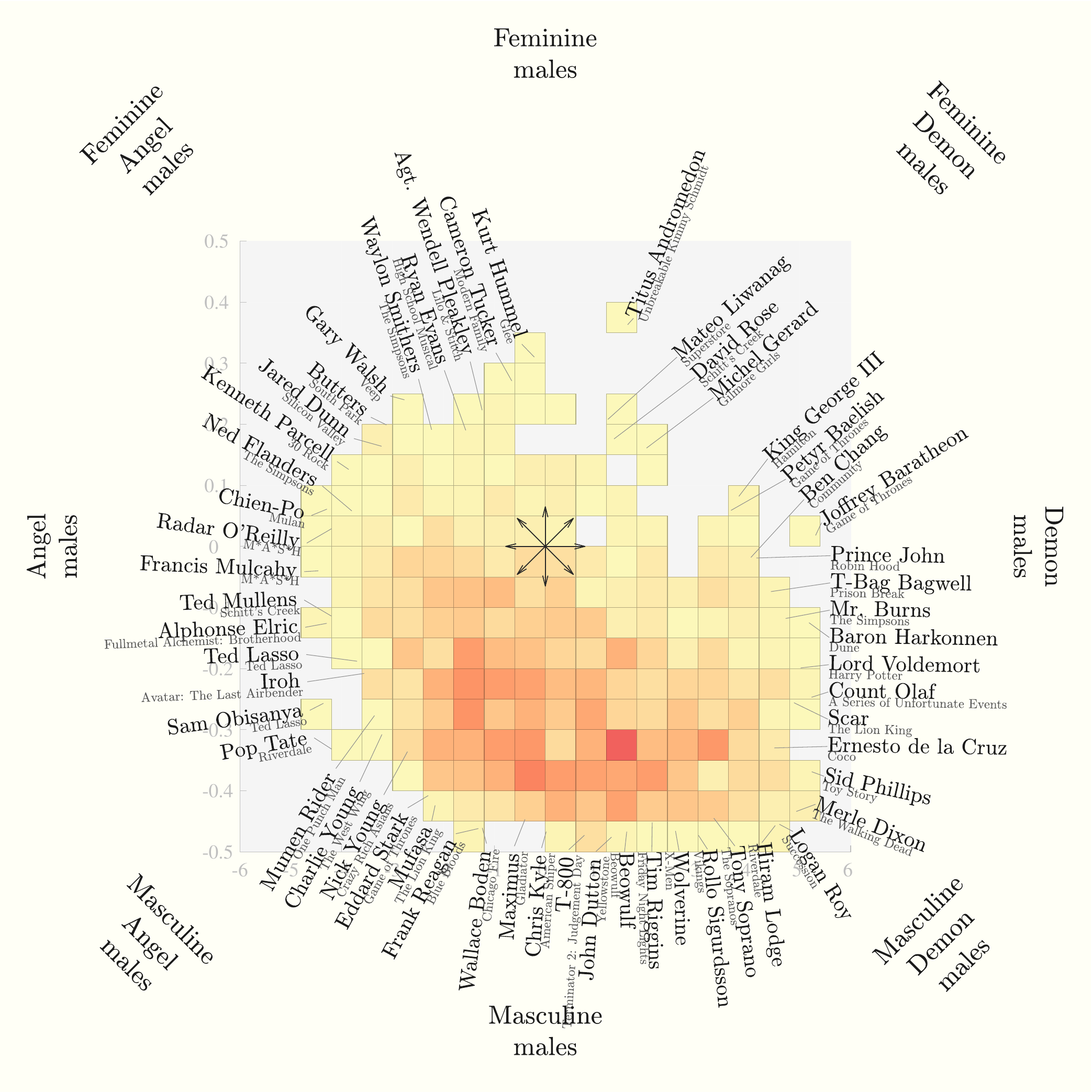}

\caption{This ousiogram depicts canonically male characters with their \protect\traitlinksimple{masculine}{feminine} score on the vertical axis and their \protect\archetypesemdiff{2} rating on the horizontal axis. The color of each block corresponds to the number of characters that demonstrates scores within the block's range; darker colors indicate more characters fall within that range than lighter color blocks, which may only contain one character.}
\label{fig:ousiogram-feminine-masculine-male-02}
\end{figure*}

\begin{figure*}[t]

\includegraphics[width=\textwidth]{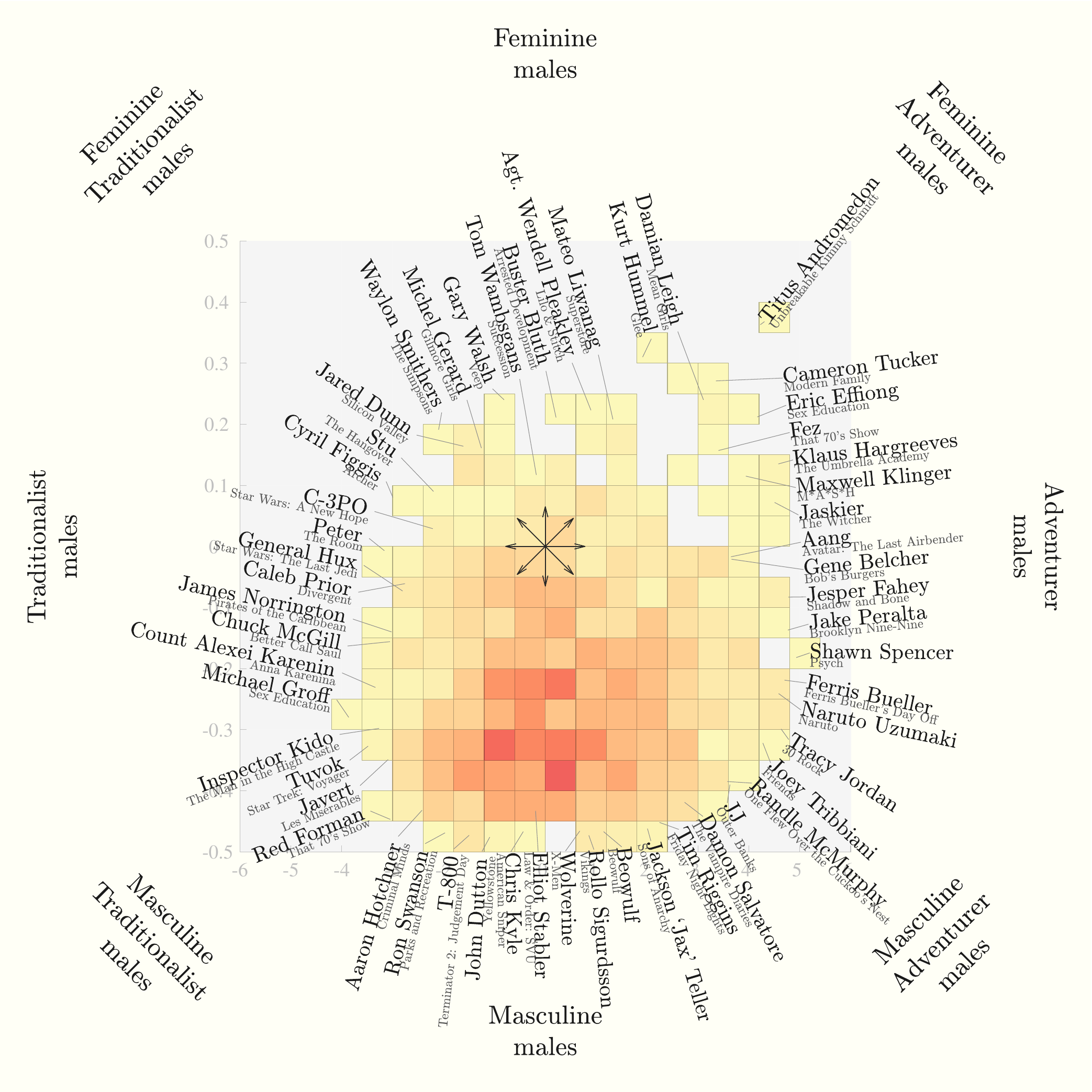}

\caption{
This ousiogram depicts canonically male characters with their \protect\traitlinksimple{masculine}{feminine} score on the vertical axis and their \protect\archetypesemdiff{3} rating on the horizontal axis. The color of each block corresponds to the number of characters that demonstrates scores within the block's range; darker colors indicate more characters fall within that range than lighter color blocks, which may only contain one character.}
\label{fig:ousiogram-feminine-masculine-male-03}
\end{figure*}

\begin{figure*}[t]

\includegraphics[width=\textwidth]{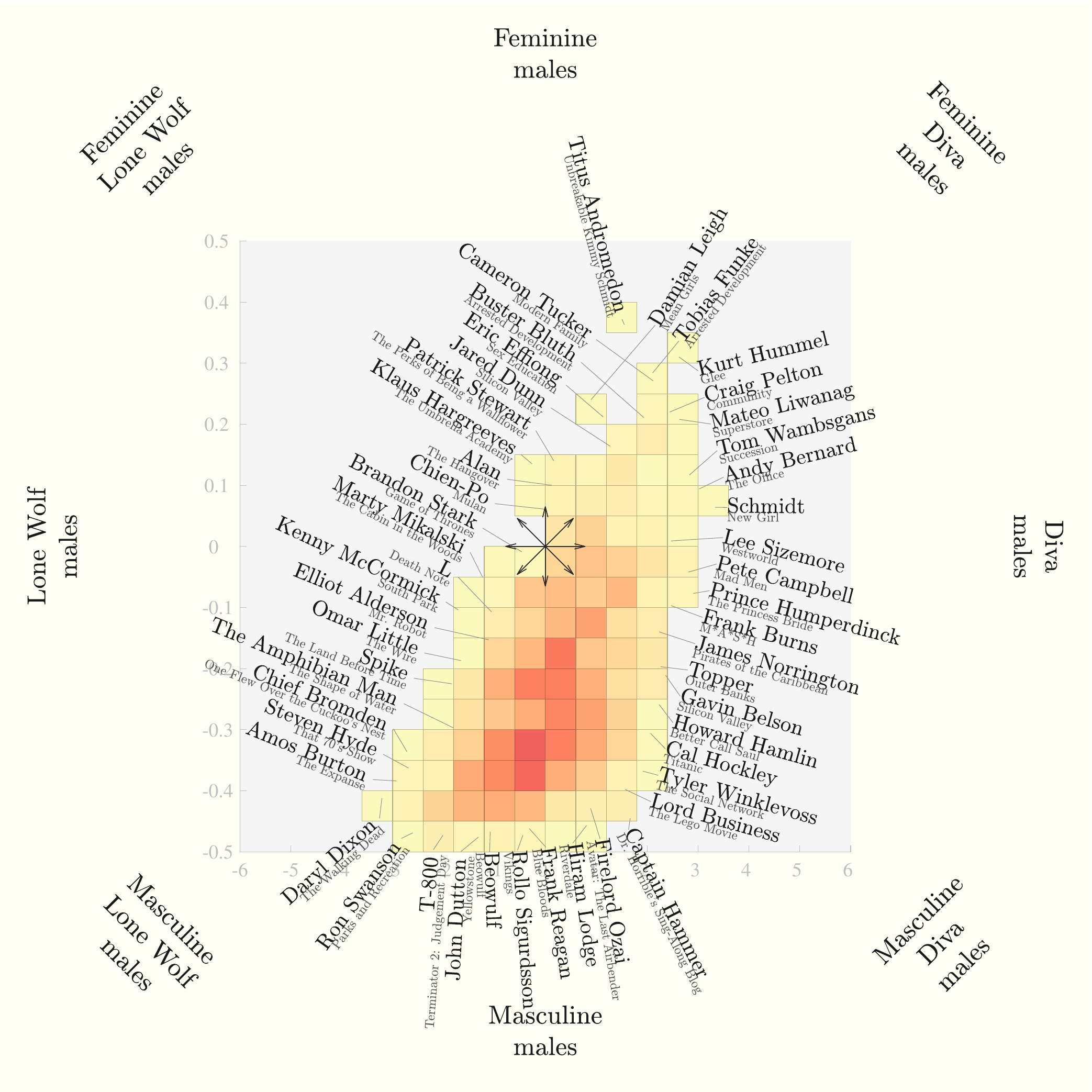}

\caption{
This ousiogram depicts canonically male characters with their \protect\traitlinksimple{masculine}{feminine} score on the vertical axis and their \protect\archetypesemdiff{4} rating on the horizontal axis. The color of each block corresponds to the number of characters that demonstrates scores within the block's range; darker colors indicate more characters fall within that range than lighter color blocks, which may only contain one character.}
\label{fig:ousiogram-feminine-masculine-male-04}
\end{figure*}

\begin{figure*}[t]

\includegraphics[width=\textwidth]{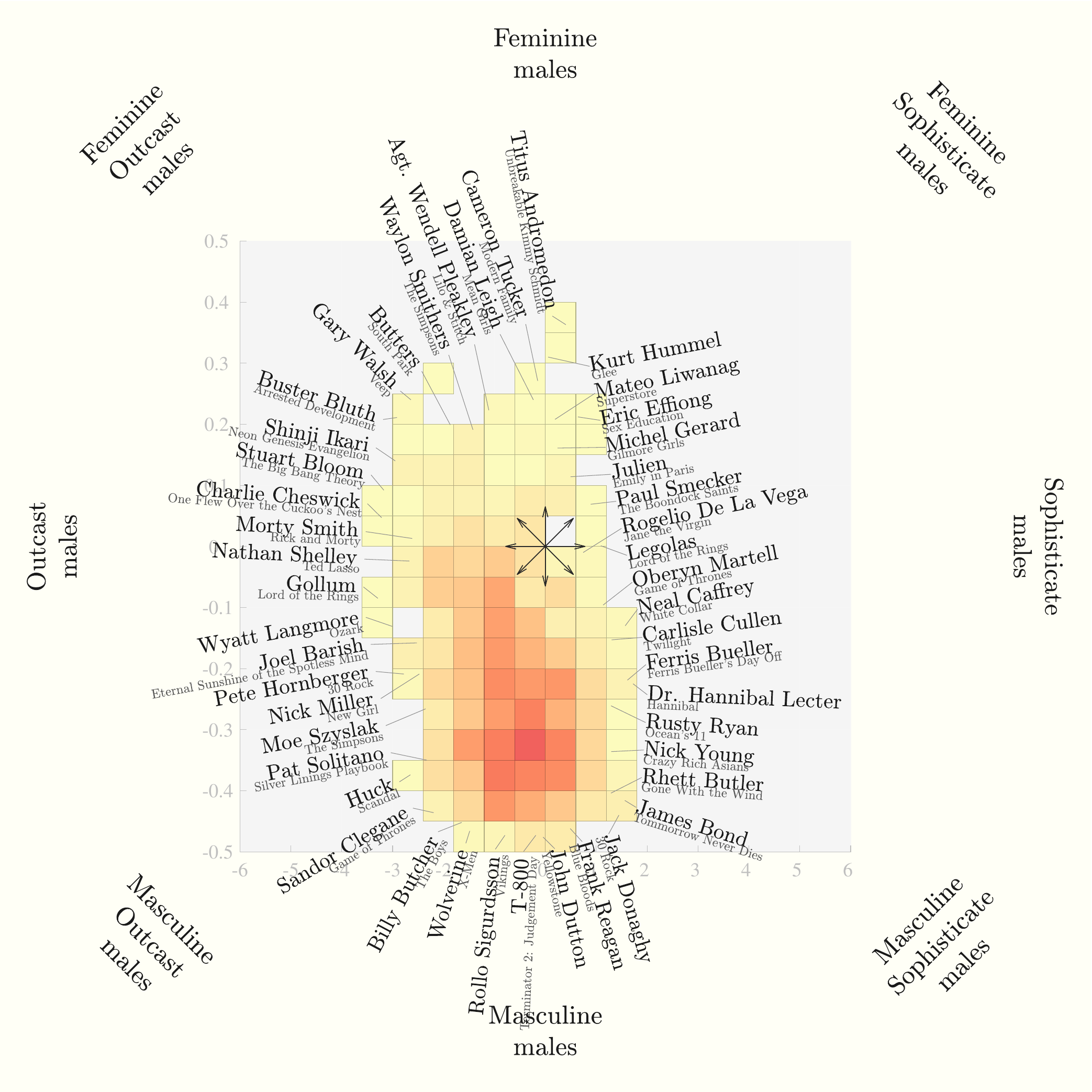}

\caption{
This ousiogram depicts canonically male characters with their \protect\traitlinksimple{masculine}{feminine} score on the vertical axis and their \protect\archetypesemdiff{5} rating on the horizontal axis. The color of each block corresponds to the number of characters that demonstrates scores within the block's range; darker colors indicate more characters fall within that range than lighter color blocks, which may only contain one character. }
\label{fig:ousiogram-feminine-masculine-male-05}
\end{figure*}

\begin{figure*}[t]

\includegraphics[width=\textwidth]{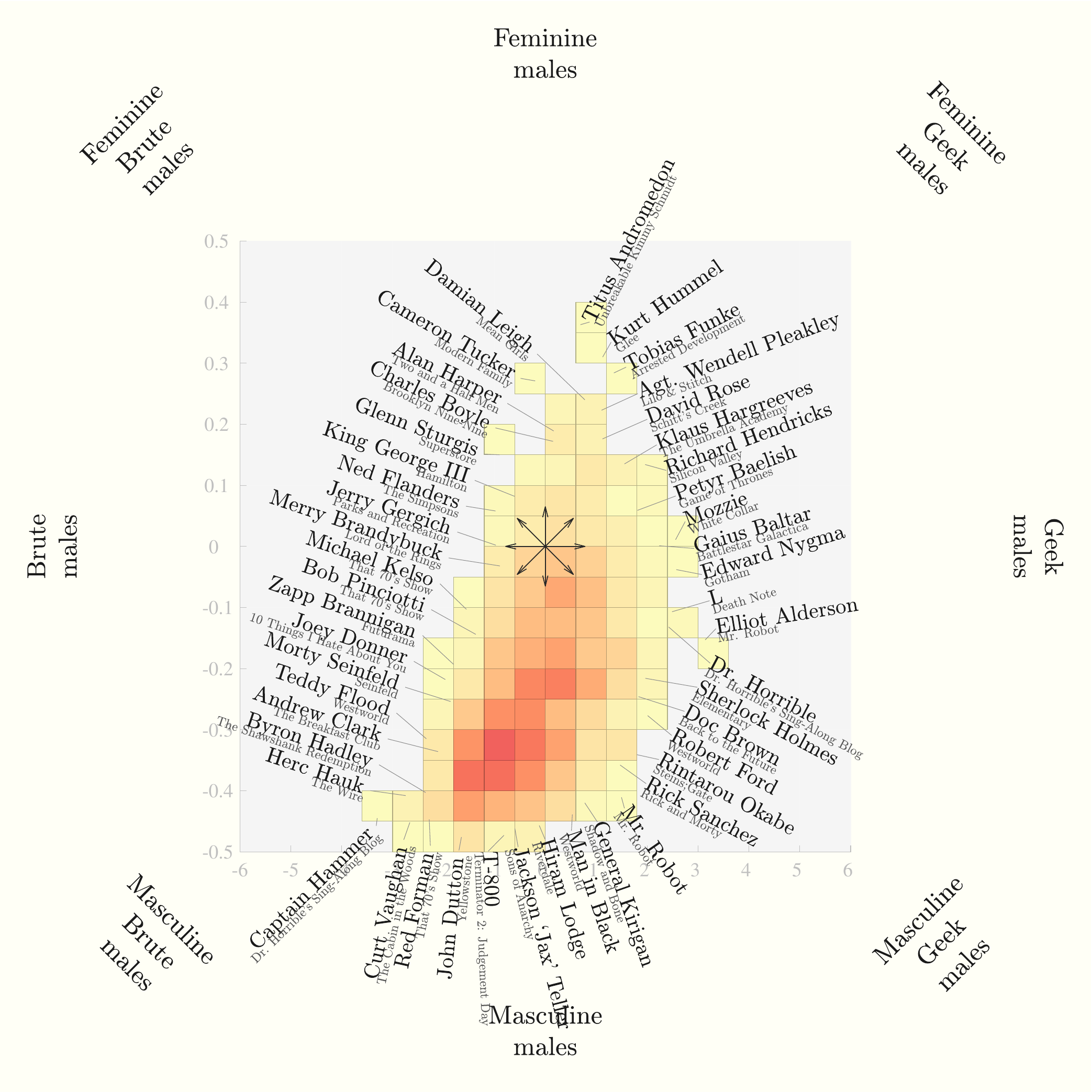}

\caption{
This ousiogram depicts canonically male characters with their \protect\traitlinksimple{masculine}{feminine} score on the vertical axis and their \protect\archetypesemdiff{6} rating on the horizontal axis. The color of each block corresponds to the number of characters that demonstrates scores within the block's range; darker colors indicate more characters fall within that range than lighter color blocks, which may only contain one character. 
}
\label{fig:ousiogram-feminine-masculine-male-06}
\end{figure*}

\begin{figure*}[t]

\includegraphics[width=\textwidth]{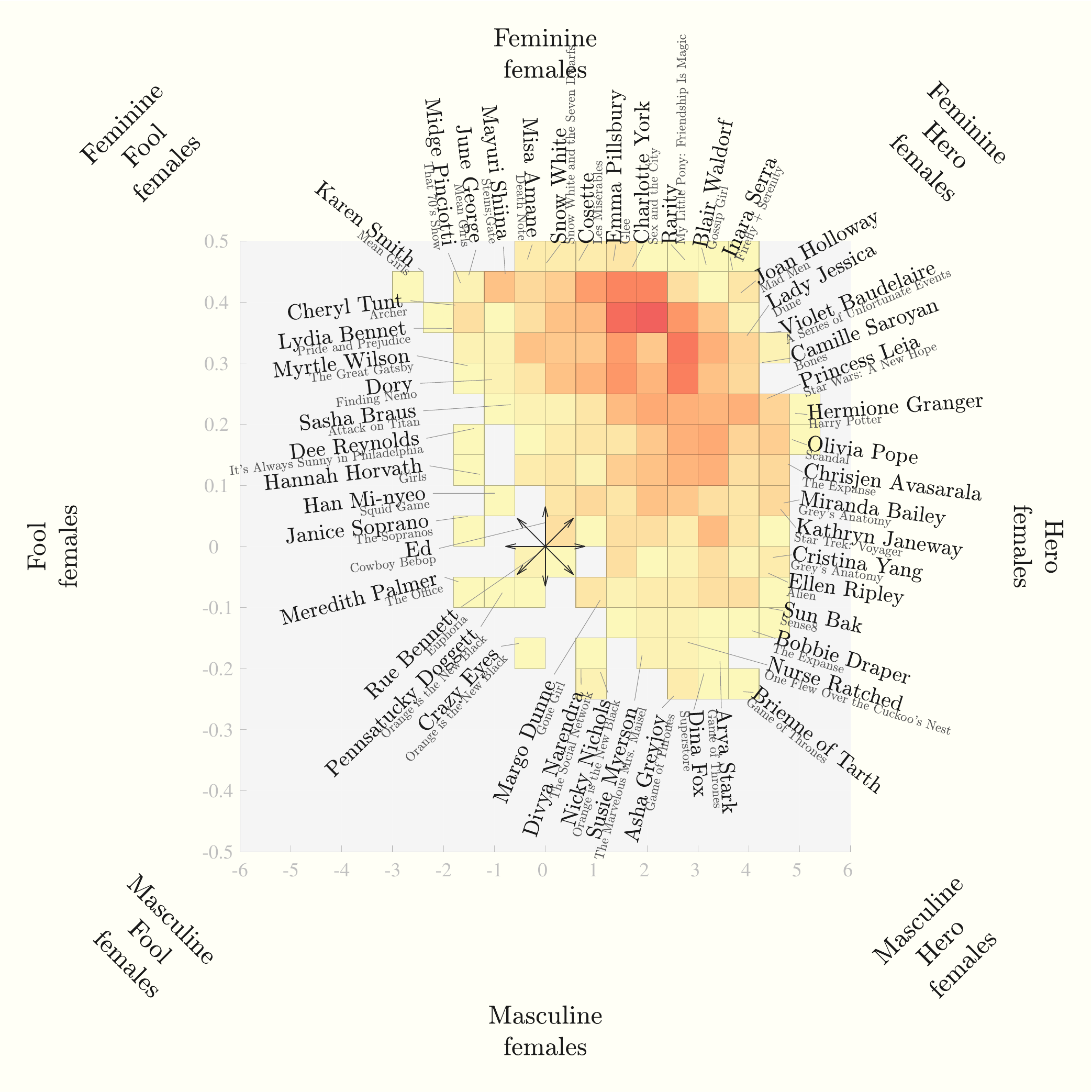}

\caption{
This ousiogram depicts canonically female characters with their \protect\traitlinksimple{masculine}{feminine} score on the vertical axis and their \protect\archetypesemdiff{1} rating on the horizontal axis. The color of each block corresponds to the number of characters that demonstrates scores within the block's range; darker colors indicate more characters fall within that range than lighter color blocks, which may only contain one character.}
\label{fig:ousiogram-feminine-masculine-female-01}
\end{figure*}

\begin{figure*}[t]

\includegraphics[width=\textwidth]{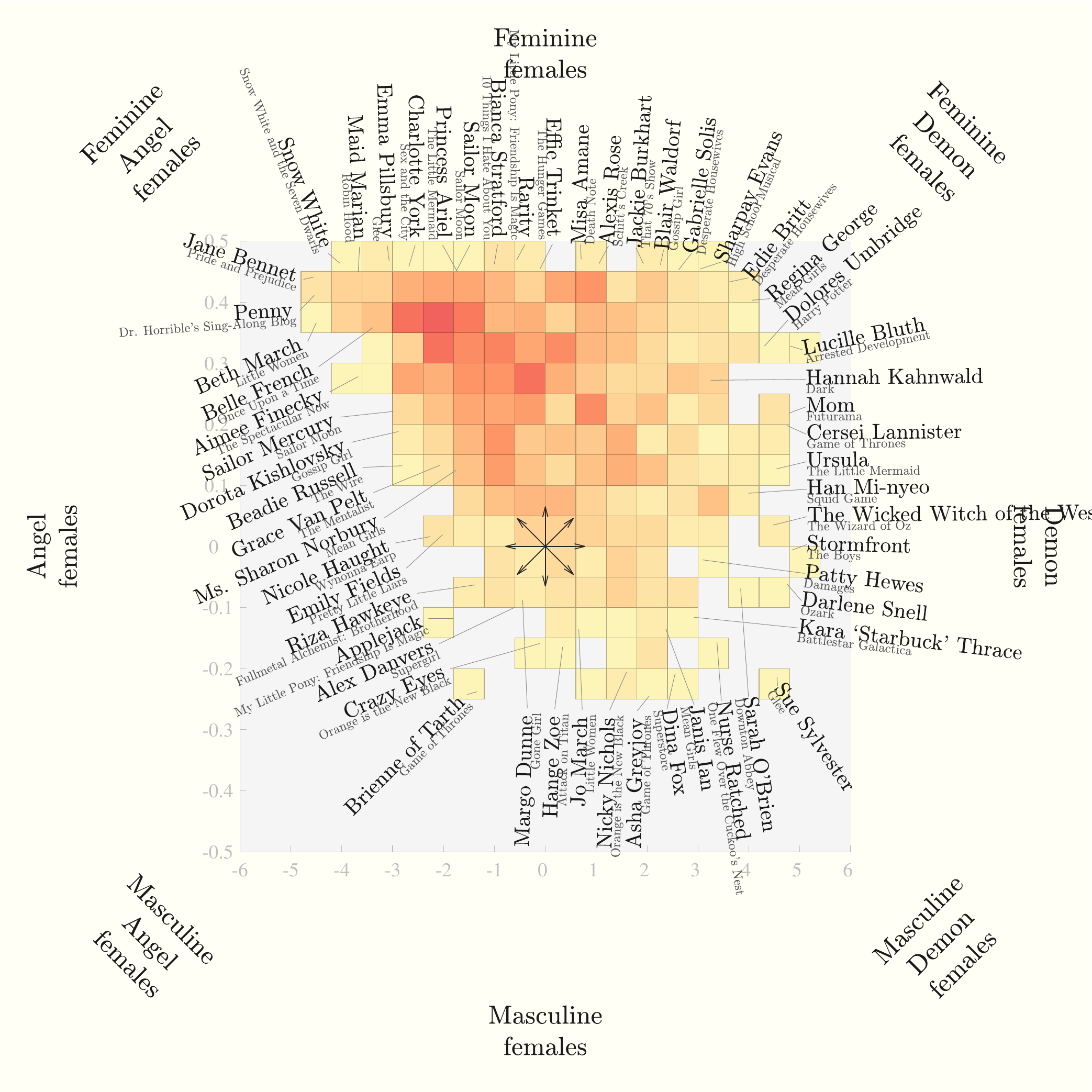}

\caption{
This ousiogram depicts canonically female characters with their \protect\traitlinksimple{masculine}{feminine} score on the vertical axis and their \protect\archetypesemdiff{2} rating on the horizontal axis. The color of each block corresponds to the number of characters that demonstrates scores within the block's range; darker colors indicate more characters fall within that range than lighter color blocks, which may only contain one character.
}
\label{fig:ousiogram-feminine-masculine-female-02}
\end{figure*}

\begin{figure*}[t]

\includegraphics[width=\textwidth]{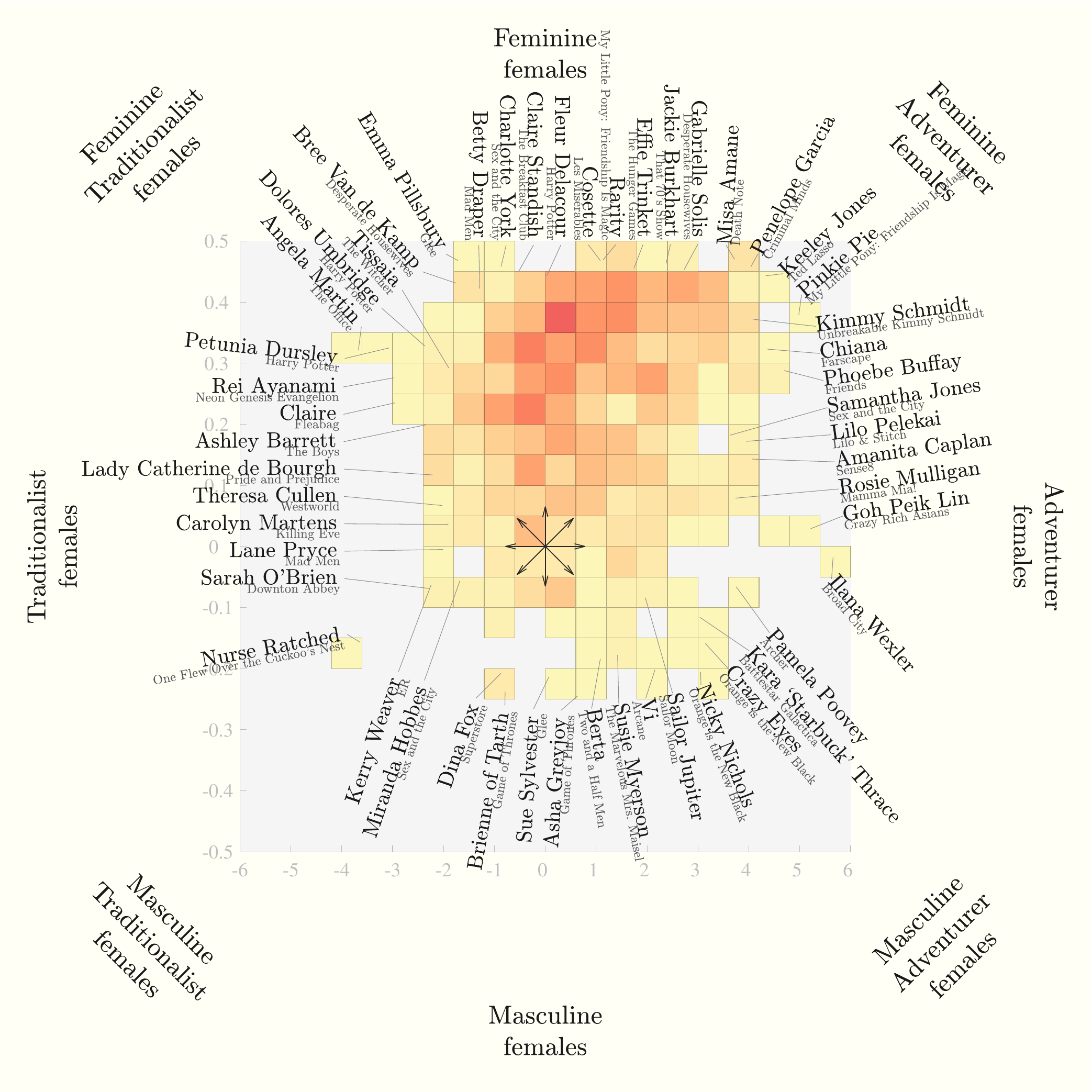}

\caption{
This ousiogram depicts canonically female characters with their \protect\traitlinksimple{masculine}{feminine} score on the vertical axis and their \protect\archetypesemdiff{3} rating on the horizontal axis. The color of each block corresponds to the number of characters that demonstrates scores within the block's range; darker colors indicate more characters fall within that range than lighter color blocks, which may only contain one character.
}
\label{fig:ousiogram-feminine-masculine-female-03}
\end{figure*}

\begin{figure*}[t]

\includegraphics[width=\textwidth]{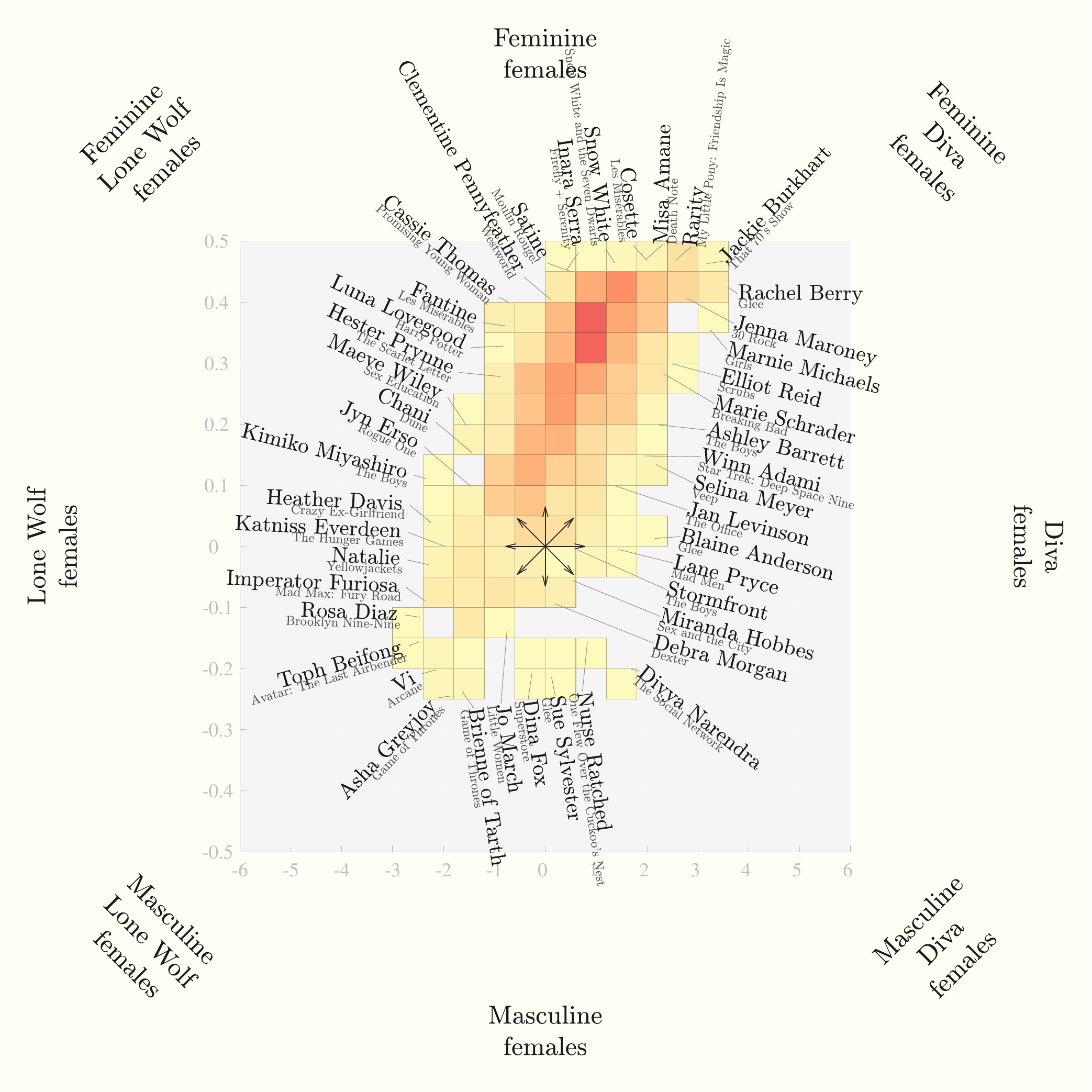}

\caption{
This ousiogram depicts canonically female characters with their \protect\traitlinksimple{masculine}{feminine} score on the vertical axis and their \protect\archetypesemdiff{4} rating on the horizontal axis. The color of each block corresponds to the number of characters that demonstrates scores within the block's range; darker colors indicate more characters fall within that range than lighter color blocks, which may only contain one character.
}
\label{fig:ousiogram-feminine-masculine-female-04}
\end{figure*}

\begin{figure*}[t]

\includegraphics[width=\textwidth]{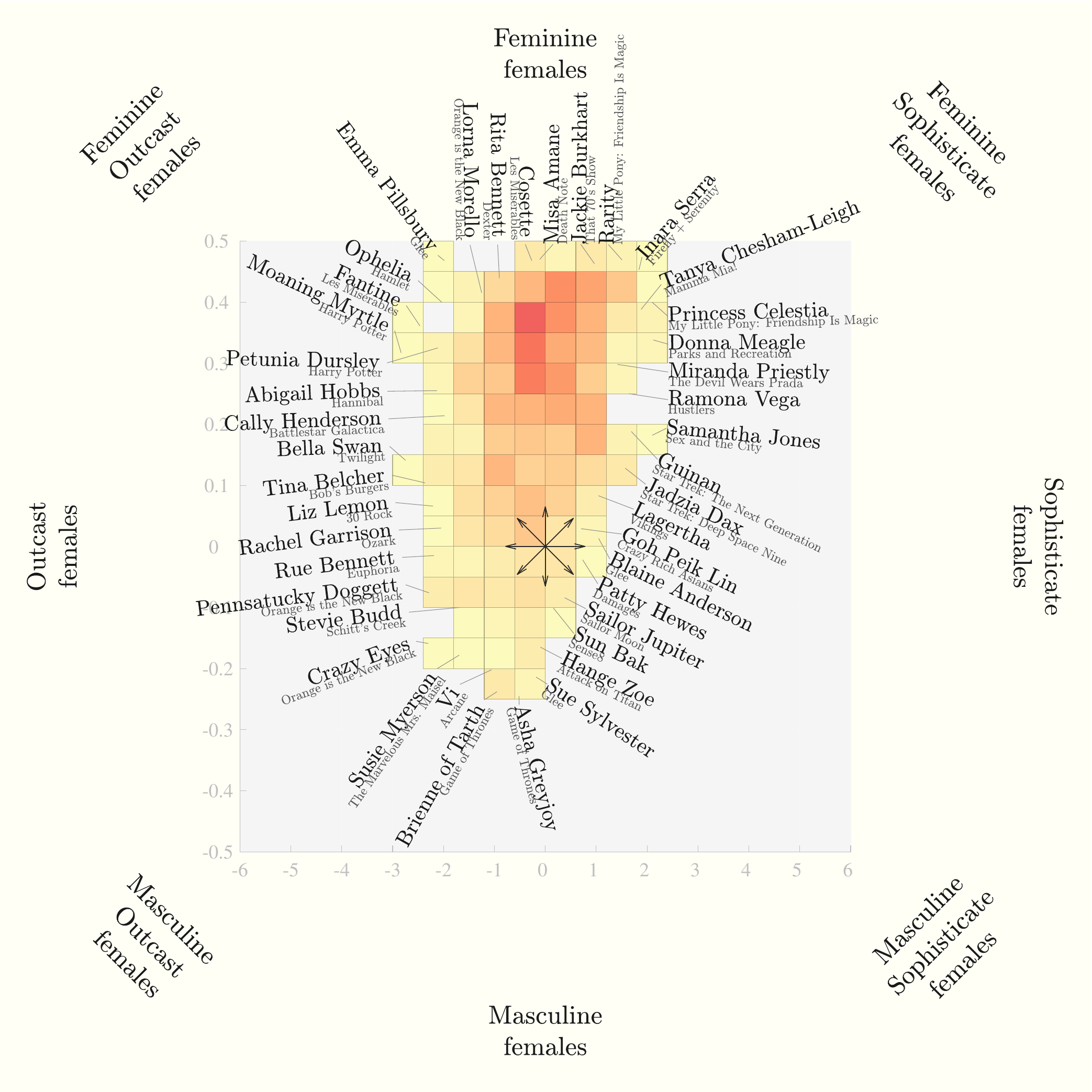}

\caption{
This ousiogram depicts canonically female characters with their \protect\traitlinksimple{masculine}{feminine} score on the vertical axis and their \protect\archetypesemdiff{5} rating on the horizontal axis. The color of each block corresponds to the number of characters that demonstrates scores within the block's range; darker colors indicate more characters fall within that range than lighter color blocks, which may only contain one character.
}
\label{fig:ousiogram-feminine-masculine-female-05}
\end{figure*}

\begin{figure*}[t]

\includegraphics[width=\textwidth]{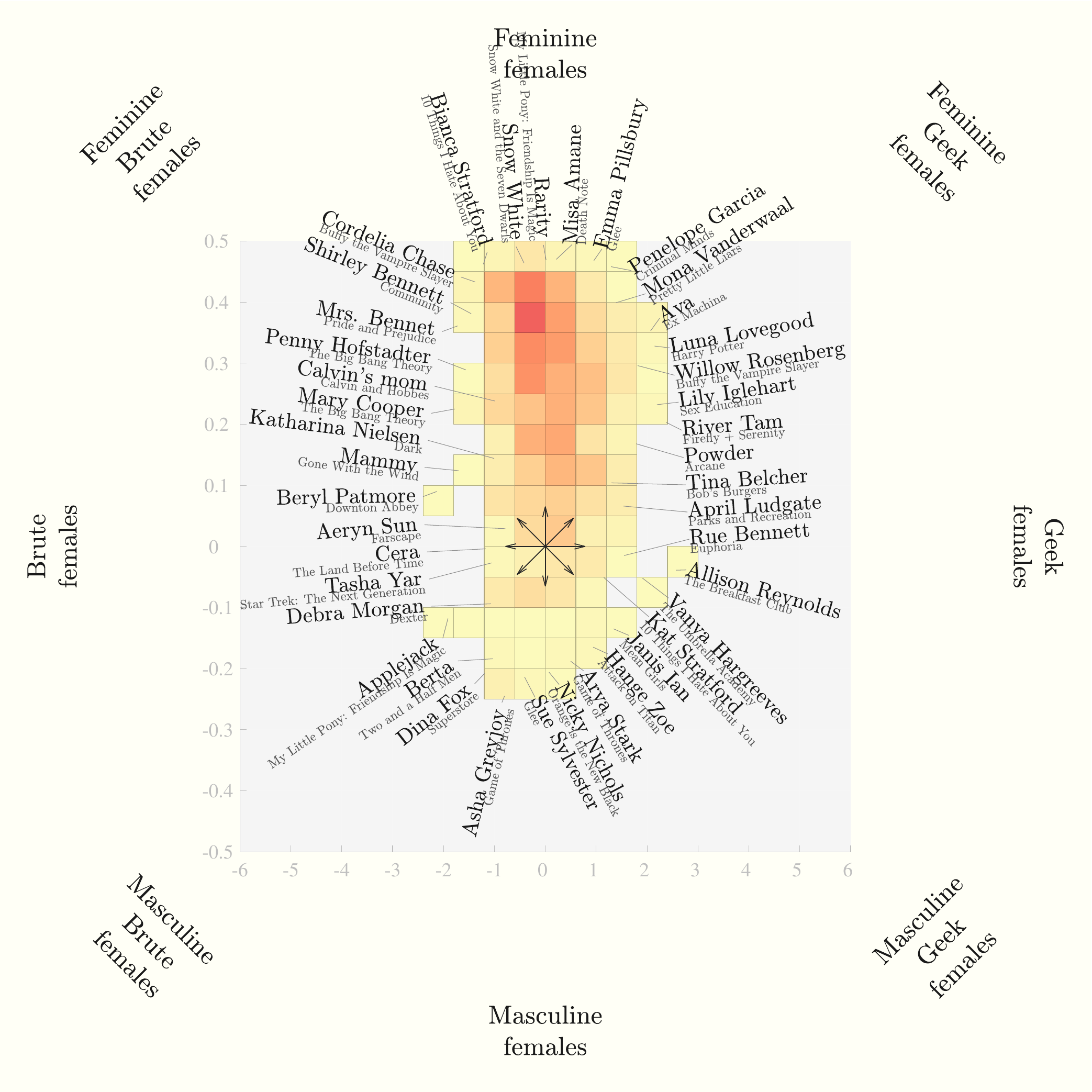}

\caption{
This ousiogram depicts canonically female characters with their \protect\traitlinksimple{masculine}{feminine} score on the vertical axis and their \protect\archetypesemdiff{6} rating on the horizontal axis. The color of each block corresponds to the number of characters that demonstrates scores within the block's range; darker colors indicate more characters fall within that range than lighter color blocks, which may only contain one character.
}
\label{fig:ousiogram-feminine-masculine-female-06}
\end{figure*}